\RequirePackage{fix-cm}
\documentclass[twocolumn,epjc3]{svjour3}
\smartqed  
\RequirePackage{graphicx}
\RequirePackage{braket}
\RequirePackage{amsmath}
\RequirePackage{amssymb}
\RequirePackage{color}
\RequirePackage{graphicx}
\RequirePackage{subfig}
\RequirePackage{hyperref}
\RequirePackage{booktabs}
\RequirePackage{tabularx}
\RequirePackage{tabulary}
\RequirePackage{adjustbox}
\RequirePackage[numbers,sort&compress]{natbib}
\RequirePackage{comment}

\usepackage[mathscr,scaled=1.15]{urwchancal}
\DeclareFontFamily{OT1}{pzc}{}
\DeclareFontShape{OT1}{pzc}{m}{it}%
{<-> s * [1.15] pzcmi7t}{}
\DeclareMathAlphabet{\mathpzc}{OT1}{pzc}{m}{it}

\usepackage{CJKutf8}

\journalname{Eur. Phys. J. C}
\begin{document}
\begin{CJK}{UTF8}{gbsn}

\title{$\,$\\[-6ex]\hspace*{\fill}{\normalsize{\sf\emph{Preprint no}.\
NJU-INP 105/25}}\\[1ex]Exclusive photoproduction of light and heavy vector mesons: thresholds to very high energies}


\author{Lin Tang (唐淋)\thanksref{addr1,addr2}%
       $\,^{\href{https://orcid.org/0009-0001-2324-9963}{\textcolor[rgb]{0.00,1.00,0.00}{\sf ID}}}$
       \and
      Hui-Yu Xing (邢惠瑜)\thanksref{addr1,addr2}%
    $\,^{\href{https://orcid.org/0000-0002-0719-7526}{\textcolor[rgb]{0.00,1.00,0.00}{\sf ID}}}$
        \and \\ Minghui Ding (丁明慧)\thanksref{e2,addr1,addr2}%
    $^{,\href{https://orcid.org/0000-0002-3690-1690}{\textcolor[rgb]{0.00,1.00,0.00}{\sf ID}}}$
        \and Craig D. Roberts\thanksref{e3,addr1,addr2}%
       $^{,\href{https://orcid.org/0000-0002-2937-1361}{\textcolor[rgb]{0.00,1.00,0.00}{\sf ID}}}$
}

\thankstext{e2}{e-mail: mhding@nju.edu.cn}
\thankstext{e3}{e-mail: cdroberts@nju.edu.cn}

\institute{School of Physics, \href{https://ror.org/01rxvg760}{Nanjing University}, Nanjing, Jiangsu 210093, China \label{addr1}
           \and
           Institute for Nonperturbative Physics, \href{https://ror.org/01rxvg760}{Nanjing University}, Nanjing, Jiangsu 210093, China \label{addr2}
}

\date{2026 February 05} 

\maketitle

\begin{abstract}
A reaction model for $\gamma + p \to V + p$, $V=\rho^0, \phi, J/\psi, \Upsilon$, which exposes the quark-antiquark content of the photon in making the transition $\gamma\to {\mathpzc q} \bar{\mathpzc q} + \mathbb P \to V$, where ${\mathpzc q}$ depends on $V$, and couples the intermediate ${\mathpzc q} \bar{\mathpzc q}$ system to the proton's valence quarks via Pomeron ($\mathbb P$) exchange, is used to deliver a unified description of available data -- both differential and total cross sections -- from near threshold to very high energies, $W$, for all the $V$-mesons.  For the $\Upsilon$, this means $10\lesssim W/{\rm GeV} \lesssim 2\,000$.  Also provided are predictions for the power-law exponents that are empirically used to characterise the large-$W$ behaviour of the total cross sections and slope parameters characterising the near-threshold differential cross sections.  Appealing to notions of vector meson dominance, the latter have been interpreted as vector-meson--proton scattering lengths.  The body of results indicates that it is premature to link any $\gamma + p \to V + p$ data with, for instance, in-proton gluon distributions, the quantum chromodynamics trace anomaly, or pentaquark production.  Further developments in reaction theory and higher precision data are required before the validity of any such links can be assessed.
\end{abstract}
\end{CJK}



\section{Introduction}
\label{sec:1}
Exclusive vector-meson photoproduction,
\begin{align}
\gamma+p\rightarrow V+p\,,\quad \text{with\,\,} V=\rho,\phi,J/\psi,\Upsilon,
\end{align}
provides a clean environment to study diffractive scattering from the proton and thereby obtain information about the reaction mechanism and both hadrons involved in the process.  (Herein, we consider only ground-state vector-mesons.)

Understanding the energy dependence of these cross sections has long challenged theory.
Contrary to perturbative quantum chromodynamics (QCD) expectations, which suggest that total cross sections should fall as the center‑of‑mass energy-squared, $s=W^2$, increases, experiments show that such photoproduction cross sections actually rise with $s$ \cite{Baksay:1978sg}.
This issue was analysed, \emph{e.g}., in Ref.\,\cite{Foldy:1963zz}, with the conclusion that the mechanism responsible should correspond to exchange of something that has isospin zero and is even under charge conjugation.
Interactions of this kind can be expressed in Regge phenomenology via Pomeron, $\mathbb P$, exchange, with a trajectory whose intercept $\alpha_{\mathbb{P}}(0)>1$.

The QCD character of the Pomeron is widely discussed \cite{Donnachie:2002en, Gribov:2003nw}.
It is typically supposed to represent the collective exchange of a family of colourless, crossing-even states, with two-gluon exchange being the leading contributor \cite{Forshaw:1997dc}.
Some phenomenological approaches distinguish between a hard (not necessarily perturbative) and soft Pomeron \cite{Donnachie:1998gm}; and connections with glueball physics are also discussed \cite{Meyer:2004jc, Boschi-Filho:2005xct}.

\begin{figure}[t]
\hspace*{-0.05\columnwidth}\includegraphics[width=1.1\columnwidth]{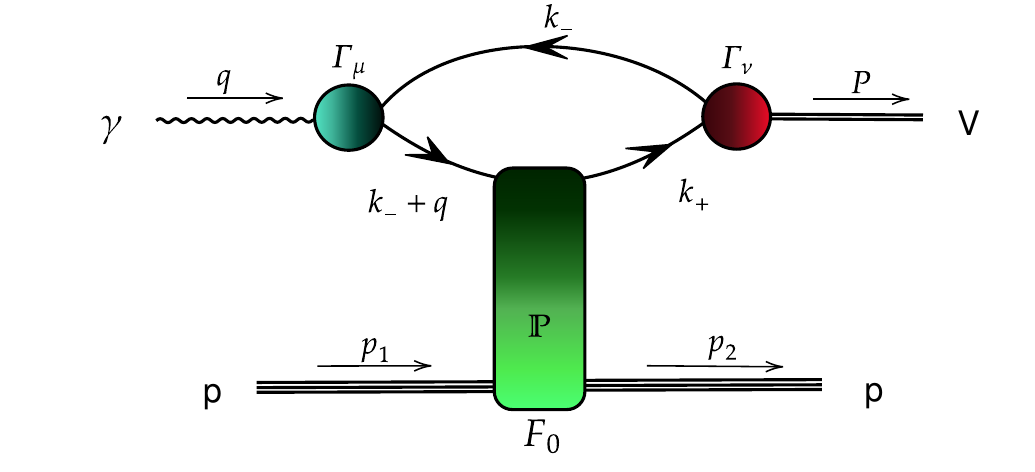}
\caption{\label{FigPomMechanism}
Reaction model for $\gamma + p \to V + p$, referred to hereafter as $\mathbb P - $dyn.
The quark-antiquark, ${\mathpzc q}\bar {\mathpzc q}$, component of the dressed-photon is probed by Pomeron exchange with the proton, producing an on-shell vector meson, $V$.
Referring to Eq.\,\eqref{EqtqP}:
the green rectangle describes Pomeron exchange, $\Gamma^{\mathbb P}$, including its couplings to the proton and quark -- see Sect.\,\ref{SecPomeron}.
Further, the solid black curve is $S_q$, the dressed $q$ quark propagator;
the shaded lighter-blue circle is $\Gamma_\mu^\gamma$, the dressed-$\gamma q \bar q$ vertex;
and the shaded red circle is $\Gamma_\nu$, the $V$-meson Bethe-Salpeter amplitude -- see Sect.\,\ref{sec:3}.
Kinematics: $s=W^2 = -(p_2+ P)^2$; $t=-(p_2-p_1)^2$.
%
}
\end{figure}

Twenty and more years ago, a variety of such models were proposed as motivations for the measurement of $\gamma+p\rightarrow V+p$ photoproduction reactions; see, \emph{e.g}., Refs.\,\cite{Nemchik:1994fp, Cudell:1996hx, Frankfurt:1998yf, Brodsky:2000zc, Forshaw:2003ki}.
A number were built upon the assumption of factorisation for hard exclusive processes, \emph{e.g}., \cite{Nemchik:1994fp, Frankfurt:1998yf, Brodsky:2000zc}, and different studies canvassed the possibilities of distinct microscopic interaction mechanisms.
Notwithstanding that, all such models attempted to draw a connection between $V$-meson photoproduction reactions and features of the target and produced meson, like in-target gluon structure functions, with such links drawn using \emph{Ans\"atze} for the light-front wave functions of the bound states involved.
Infrared freezing of the QCD running coupling was also sometimes assumed in the interaction models \cite{Nemchik:1994fp, Cudell:1996hx}, although with saturation values chosen which are significantly less than that associated with QCD's process-indepen\-dent effective charge \cite{Cui:2019dwv, Deur:2023dzc, Brodsky:2024zev}.
Owing to their formulation assumptions, the models were only considered plausible at energies far above threshold.
As will become clear, our framework represents a significant departure from such approaches.
For instance, \emph{inter alia}:
no call is made to factorisation for hard exclusive processes;
the required bound-state wave functions and constituent propagators are consistently calculated using an interaction that derives from modern analyses of QCD's gauge sector;
and predictions are simultaneously made for all $V$-meson final states, at energies that range, in each case, from threshold to very far above.

Continuing with this last point, having noted that the high-energy behaviour of $\gamma+p\rightarrow V+p$ is described by $\mathbb P$ exchange, it is natural, as herein, to seek an explanation for the cross section at low-energies and determine whether a unified approach to the entire $s$-do\-main is possible.
Both these issues were addressed in Ref.\,\cite{Pichowsky:1996tn}, which explored the reaction mechanism depicted in Fig.\,\ref{FigPomMechanism} using phenomenological inputs for the required elements.
This approach sees the ${\mathpzc q}\bar {\mathpzc q}$ component of the photon, where $\mathpzc q$ is chosen from the set $\{{\mathpzc q}=u/d,s,c,b\}$, depending on $V$, exposed by interaction with the proton via $\mathbb P$-exchange and then transformed into the on-shell vector meson when sufficient reaction energy is available, \emph{i.e}., once the threshold energy is exceeded:
\begin{equation}
W_{\rm th}^V = m_p + m_V\,.
\end{equation}
Here $m_p=0.939\,$GeV is the proton mass and $m_V$ is that of the vector meson produced.

The $\gamma\to {\mathpzc q}\bar {\mathpzc q} +\mathbb{P}\to V$ transition matrix element in Fig.\,\ref{FigPomMechanism} can be calculated using QCD propagators and vertices, which may be obtained using either continuum or lattice Schwinger function methods (C[L]SMs).
There are three nonperturbative, matrix-valued Schwin\-ger functions; and in the Ref.\,\cite{Tang:2024pky} study of $\gamma + p \to J/\psi + p$, they were calculated using CSMs \cite{Eichmann:2016yit, Burkert:2017djo, Roberts:2021nhw, Binosi:2022djx, Ding:2022ows, Ferreira:2023fva, Achenbach:2025kfx}.
The analysis delivered parameter-free predictions for both differential and total cross sections from near threshold to very high energies that are in agreement with much available data.

Building further upon Refs.\,\cite{Pichowsky:1996tn, Tang:2024pky}, it is relevant to determine whether the reaction mechanism indicated in Fig.\,\ref{FigPomMechanism}, with elements calculated using CSMs, can deliver realistic, parameter-free predictions for photoproduction reactions involving other vector mesons, both heavier and lighter than the $J/\psi$.
The associated comparisons with results from decades of measurements of $\rho$, $\phi$, $J/\psi$, $\Upsilon$ photoproduction reactions, both differential and total cross sections, over a huge range of energies, would provide valuable information on $\mathbb P$-exchange universality and other features of photoproduction processes.
Notably, spurred by interest in the proton's gluon content and possible pentaquark states \cite{Chen:2020ijn, Anderle:2021wcy, AbdulKhalek:2021gbh}, the experimental data set has recently been expanded by experiments at Jefferson Lab \cite[GlueX]{GlueX:2023pev} and CERN \cite[LHCb]{LHCb:2024pcz}.


This discussion is arranged as follows.
Section~\ref{sec:2} presents the reaction model and our Pomeron exchange framework for vector meson photoproduction.
It is followed, in Sect.\,\ref{sec:3}, by an explanation of the CSM calculation of the Schwinger functions required to obtain the $\gamma + p \to V + p$ matrix element sketched in Fig.\,\ref{FigPomMechanism}.  That matrix element enables completion of the cross section calculations.
Section~\ref{sec:4} explains how one may infer phenomenological vector-meson--proton scattering lengths from near-threshold differential cross section data using notions of vector meson dominance (VMD).
Whilst not rigorous, these scattering lengths can be useful in constraining model studies.
They even become objective comparative measures after reinterpretation as a near-threshold slope parameter for each cross section.
A large collection of numerical results is reported in Sect.\,\ref{sec:5} along with relevant data comparisons.
That discussion continues in Sect.\,\ref{SecPower}, which analyses the large-$W$ behaviour of photoproduction total cross sections.
Numerical results for the near-threshold slope parameter (VMD-based $V-p$ scattering length) are reported in Sect.\,\ref{SecSlope}.
Section~\ref{sec:6} provides a summary and perspective.

\section{Reaction model and Pomeron exchange}
\label{sec:2}
\subsection{Dynamical photon to meson transition}
\label{sec:21}
The image in Fig.\,\ref{FigPomMechanism} corresponds to the following matrix element \cite{Pichowsky:1996tn, Donnachie:1984xq, Donnachie:1987pu}:
\begin{align}
{\cal I}_\mu^V(W,t) & = \langle V(P;\lambda) p(p_2) | \bar {\mathpzc q} \gamma_\mu {\mathpzc q}| p(p_1)\rangle  \nonumber \\
& =
2 t_{\mu \alpha \nu}^V(q,P) \epsilon_\nu^\lambda(P) \bar u(p_2) \tilde G_\alpha(w^2,t) u(p_1)\,,
\label{eq:1}
\end{align}
where
$\epsilon_\nu^\lambda(P)$ is the $V$-meson polarisation vector,
$u(p_1)$, $\bar u(p_2)$ are spinors for the incoming and outgoing proton, \linebreak
$\tilde G_\alpha(w^2,t) $ represents the quark-Pomeron-proton interaction;
and the factor ``2'' expresses the equivalence between striking the upper and lower valence quark lines in the quark loop.
The invariant mass of the $\gamma p$ system is $s=W^2=-(q+p_{1})^{2}$ and the momentum transfer is $t=-(p_{1}-p_{2})^{2}$.

The exposed quark loop integration in Fig.\,\ref{FigPomMechanism} receives most support on the neighbourhood $k^2 \simeq 0$, \emph{i.e}., on the domain of near-zero relative momentum within the final state meson, whereat the meson's Bethe-Salpeter amplitude is peaked.
One may therefore write $w^2=-(q-P/2+p_1)^2$.
In this case, the essentially dynamical component of the reaction model, \emph{viz}.\ the
$\gamma\to {\mathpzc q} \bar {\mathpzc q} + \mathbb P \to V$ transition matrix element, can be written as follows:
\begin{align}
& t_{\mu \alpha \nu}^V(q,P)  =  e_V N_c
{\rm tr}_D \int \frac{d^4 k}{(2\pi)^4}
S_{\mathpzc q}(k_-) \Gamma_\mu^\gamma(k_-,k_-+q) \nonumber \\
& \quad \times S_{\mathpzc q}(k_-+q) \Gamma^{\mathbb P}_\alpha(k_-+q,k_+) S_{\mathpzc q}(k_+) \Gamma_\nu^V(k_+,k_-)\,,
\label{EqtqP}
\end{align}
where
$N_c=3$;
$e_{\rho^0,\phi,J/\psi,\Upsilon} = e \{1/\surd 2 , 1/3 , 2/3, 1/3\}$,
the fine-structure constant $\alpha = e^2/[4\pi]$;
the trace is over spinor indices;
$k_\pm = k \pm P/2$;
$\Gamma_\nu^V$ is the Bethe-Salpeter amplitude for an on-shell $V$ meson
 -- see, \emph{e.g}., Ref.\,\cite{Xu:2021mju};
$S_{\mathpzc q}$ is the dressed $q$ quark propagator;
 -- see, \emph{e.g}., Ref.\,\cite[Sect.\,2.C]{Roberts:2021nhw};
and $\Gamma_\mu^\gamma$ is the associated dressed photon-quark vertex.
Since we are focusing on photoproduction, then, with complete generality, one may use the Ball-Chiu form for $\Gamma_\mu^\gamma$ \cite{Ball:1980ay}:
\begin{align}
i \Gamma_\mu^{\rm BC}&(k_1,k_2) = \Sigma_A(k_1^2,k_2^2)) i \gamma_\mu 
+ (k_1+k_2)_\mu \nonumber \\
& 
\times [ \Delta_A(k_1^2,k_2^2) i \gamma\cdot(k_1+k_2)+\Delta_B(k_1^2,k_2^2) ]\,,
\label{BCV}
\end{align}
where $ \Sigma_F(k_1^2,k_2^2)=[F(k_1)^2+F(k_2^2)]/2$,
$ \Delta_F(k_1^2,k_2^2))=[F(k_1)^2-F(k_2^2)]/[k_1^2-k_2^2]$,
$F=A, B$.  
This \emph{Ansatz} is complete for photoproduction because, with $Q^2=0$, it delivers the Ward identity result:
$i \Gamma_\mu^{\rm BC}(k_-,k_-)=\partial _\mu S(k_-)$.

The remaining element in Eq.\,\eqref{EqtqP} is $ \Gamma^{\mathbb P}_\alpha$, the Pomeron-quark vertex.
Following Refs.\,\cite{Donnachie:1984xq, Pichowsky:1996tn} and for the reasons detailed therein, we employ $\Gamma^{\mathbb P}_\alpha = \gamma_\alpha \beta_{\mathpzc q}$,
where $\beta_{\mathpzc q}$ is a constant strength factor and $\gamma_\alpha$ is a standard Dirac matrix.
%

As a means of testing the sensitivity of our results to the form of $\Gamma^{\mathbb P}_\alpha(k_-+q,k_+)$, we replaced $\gamma_\alpha$ by the BC vector-vertex \emph{Ansatz}, Eq.\,\eqref{BCV}, which, evidently, involves three tensor structures (one additional Dirac vector and also a Dirac scalar) with the strength of each term determined by the associated quark propagator, $S_{\mathpzc q}$, dressing functions.
For light quarks, referring, for instance, to Ref.\,\cite[Fig.\,3]{Bhagwat:2003vw}, the BC vertex is significantly different from the bare vector vertex on $0<q^2\lesssim 4 m_p^2$.
Moreover, consistent with the character of the associated inhomogeneous Bethe-Salpeter vertex-dressing equation, the differences diminish with increasing quark current mass and, independent of quark current mass, 
$\Gamma_\alpha^{\rm BC}(k_-+q,k_+) \approx \gamma_\alpha$ for $q^2 \gg m_p^2$.
This last feature is an important constraint on our study; namely, following the arguments in Ref.\,\cite{Donnachie:1984xq}, we require that the Pomeron + quark coupling be pointlike at deep ultraviolet momenta.
(These constraints entail that any other of the commonly discussed vector-vertex \emph{Ans\"atze} and Bethe-Salpeter equation solutions would be equivalent; see, \emph{e.g}., Refs.\,\cite{Curtis:1990zs, Qin:2013mta}.)
Considering each photoproduction reaction herein, within line width, the resulting BC vertex predictions for differential and total cross sections are indistinguishable from those obtained when using $\gamma_\alpha$ alone after shifting $\beta_{\mathpzc q}\to (1-\delta_{\mathpzc q}^{BC}) \beta_{\mathpzc q}$, with $\delta_{\mathpzc q}^{BC} \lesssim 0.25$.
The associated quantitative statement is that the ${\mathpzc L}_1$ relative difference between the two sets of results is never more than 2.5\%, whether one considers the near-threshold or entire $W$ domain, or the entire $|t|$ domain of coverage for a given experiment.
(The ${\mathpzc L}_1$ relative difference between two curves is the integrated absolute-value of the difference between the two curves divided by half the integrated sum.)

Working in the center of three-momentum frame of the vector-meson--proton system, the relevant four-momenta are:
\begin{subequations}
\label{Eqmomenta}
\begin{align}
q^{\mu}&=(\vec{q},i|\vec{q}|), \\
    P^{\mu}&=(\vec{P},i\sqrt{|\vec{P}|^{2}+m_{V}^{2}}), \\
    p_{1}^{\mu}&=(-\vec{q},i\sqrt{|\vec{q}|^{2}+m_{p}^{2}}), \\
    p^{\mu}_{2}&=(-\vec{P},i\sqrt{|\vec{P}|^{2}+m_{p}^{2}})\,,
\end{align}
\end{subequations}
where the metric has uniform positive signature;
$\vec{q}$ is the three-momentum of the on-shell photon;
$\vec{P}$ is the vector-meson three-momentum;
and
\begin{equation}
|t|_{\rm min} = 2 |\vec{q}| (|\vec{P}|^2+m_{V}^2)^{1/2}-m_{V}^2 - 2 |\vec{q}| |\vec{P}|\,.
\label{Eqtmin}
\end{equation}

The differential cross section for exclusive vector meson photoproduction can now be expressed:
\begin{equation}
    \frac{d\sigma}{d\Omega}=\frac{1}{4\pi^{2}}\frac{m_{p}}{4W}\frac{|\vec{P}|}{k_{W}}
    \;     \tfrac{1}{2} \rule{-1ex}{0ex} \sum_{\text{proton spin}}(|{\cal I}_1|^{2}+|{\cal I}_2|^{2})\,,\label{eq:7}
    \end{equation}
where
\begin{equation}
k_{W}=(W^{2}-m_{p}^{2})/(2m_{p})
\label{eqkW}
\end{equation}
and $d\Omega$ is the differential solid-angle element defined by $\angle \vec{q}  \vec P$.
The corresponding differential cross section with respect to the momentum transfer is given by
    \begin{equation}
    \frac{d\sigma}{d(-t)}=\frac{\pi}{|\vec{q}||\vec{P}|}\frac{d\sigma}{d\Omega}\,.
\end{equation}
The longitudinal cross section vanishes in photoproduction.

From the expressions sketched above, it is evident that the reaction model, combined with the Pomeron-exchange framework, offers a straightforward approach to studying vector meson photoproduction. The method is computationally economical and has been successfully applied to a broad class of processes; see, \emph{e.g}., Ref.\,\cite{Donnachie:1992ny}.

\subsection{Pomeron in the reaction model}
\label{SecPomeron}
The reaction model involves two Pomeron elements.  First, the Pomeron-proton content of the coupling is \cite{Pichowsky:1996tn, Donnachie:1984xq, Donnachie:1987pu}:
\begin{equation}
\label{EqGP}
\tilde G_\alpha(w^2,t) = \gamma_\alpha F_0(t)  3 \beta_{\ell} G_{\mathbb P}(w^2,t) \,,
\end{equation}
where
``$3$'' enumerates the number of light valence de\-grees-of-freedom in the proton;
$\beta_\ell$ is the strength of the associated Pomeron + light-quark coupling;
and $F_0(t)$ is the empirical nucleon isoscalar elastic electromagnetic form factor.
This last factor is well parametrised as follows on the spacelike low-$|t|$ domain relevant herein:
\begin{equation}
F_0(t) = \frac{1-2.8 t/[4m_p^2]}{1-t/[4m_p^2]} \frac{1}{(1-t/t_0)^2}\,,
\label{EqF0t}
\end{equation}
where $t_0 = 0.71\,$GeV$^2$.

The second element in Eq.\,\eqref{EqGP} is  typically called the Pomeron propagator:
\begin{equation}
G_{\mathbb P}(s,t) = \left[\frac{s}{s_0}\right]^{\alpha_{\mathbb P}(t) - 1}
\exp\left[ - i \tfrac{\pi}{2} (\alpha_{\mathbb P}(t) - 1) \right]\,,
\label{PomProp}
\end{equation}
with
$\alpha_{\mathbb P}(t) = \alpha_0 + \alpha_1 t$,
$s_0 = 1/\alpha_1$,
and $\alpha_{0,1}$ fitted to selected small-$|t|$, large-$W$ meson photoproduction data within the context of Regge phenomenology: $\alpha_0$ is fixed by the energy dependence of the total cross section and
$\alpha_1$ by the $t$-slope of the differential cross section -- see, \emph{e.g}., Refs.\,\cite{Collins:1977jy, Irving:1977ea}.

As observed, \emph{e.g}., in Refs.\,\cite{Donnachie:1998gm, Pichowsky:1996tn, Lee:2022ymp}, data suggests that in comparing photoproduction reactions involving light or heavy vector-meson final states, different values of the pair $(\alpha_0,\alpha_1)$ are required; see, also, below.
This observation encouraged some practitioners to suggest the existence of two Pomerons \cite{Donnachie:1998gm}, one hard and one soft, with both emerging as the result of some (unknown) nonperturbative dynamics.
Given the current QCD understanding of Pomeron phenomenology and structure, we elect not to express an opinion on this matter.
Instead, in the reaction model advocated herein, there is a single (soft) exchanged object.
The object fits within Pomeron-based schemes because, amongst other things, it provides a nonlocal color-singlet exchange that produces cross sections which rise with energy.
This Pomeron represents essentially nonperturbative aspects of reactions involving hadrons; and the proton + Pomeron + quark interaction possesses an energy dependence that differs between light-quark meson and heavy-quark meson final states.
It is, perhaps, worth noting that scale-dependent Pomeron trajectory parameters may be natural in some approaches, \emph{e.g}., models based on AdS/CFT duality \cite{Dosch:2015oha}.



\begin{table*}[t]
\centering
\caption{
All parameters used herein.
Quark current-masses and parameters characterising the CSM interaction kernels for the mesons considered herein.
The quark current masses are quoted at a renormalisation scale $\zeta=\zeta_{19} = 19\,$GeV, \emph{i.e}., in the far ultraviolet so as to avoid truncation ambiguity.
The $u,d$ values were chosen to reproduce $m_\pi$, $f_\pi$;
the kaon mass was used to fix $m_s$;
$m_c$ was chosen to yield the $\eta_c$ mass;
and the $\eta_b$ mass was used to fix $m_b$ \cite{Qin:2018dqp}.
Using one-loop evolution, the listed light quark masses correspond to $\zeta = \zeta_2 = 2\,$GeV values of
$m_{\ell}^{\zeta_2} = 0.0048\,$GeV, $m_s^{\zeta_2} = 0.117\,$GeV; and regarding the heavy quarks, the listed values equate to the following Euclidean constituent quark masses \cite{Hecht:2000xa},
$M_c = 1.41\,$GeV, $M_b=4.34\,$GeV.  
All four masses correspond to those typically quoted in connection with the given flavour and are a fair match with contemporary inferences \cite{ParticleDataGroup:2024cfk}.
The same Pomeron trajectory is used for both $\rho^{0}$ and $\phi$ photoproduction \cite[Sec.\,III.A]{Pichowsky:1996tn}, and another for $J/\psi$ and $\Upsilon$ \cite[ZEUS]{ZEUS:2002wfj}.
Pomeron-quark couplings, $\beta_{\mathpzc q}$, for both reaction models considered herein -- Sects.\,\ref{sec:21}, \ref{sec:23} -- are determined as described in Sect.\,\ref{sec:24}.
\label{tab:1}}
\begin{tabular}{l|lllll|ll|l|l}
\hline
Meson & \multicolumn{5}{|c|}{Gluon model} & \multicolumn{2}{|c|}{$\mathbb{P}$ trajectory} & $\mathbb{P}-$dyn  & $\mathbb{P}-$am \\
\hline
 &$m_{\zeta_{19}}$ [GeV] & $D\omega$ [$\text{GeV}^{3}$] & $\omega$ [GeV] & $N_{f}$ & $\Lambda_{\text{QCD}}$ [GeV] & $\alpha_{0}$ & $\alpha_{1}$ [$\text{GeV}^{-2}$]  & $\beta_{\mathpzc q}$ [$\text{GeV}^{-1}$]	&$\beta_{\mathpzc q}$ [$\text{GeV}^{-1}$]\\
\noalign{\smallskip}\hline\noalign{\smallskip}
$\rho^{0}_\ell$ & 0.0034	& $0.8^{3}$ 	& 0.5  & 4 & 0.234 & 1.1 	& 0.33  & 3.5 &1.96\\
$\phi_s$ & 0.083	& $0.8^{3}$ 	& 0.5  & 4 & 0.234   & 1.1   & 0.33 	& 1.89 &1.51\\
$J/\psi_c$ &0.89	& $0.6^{3}$ 	& 0.8  & 4 & 0.234   & 1.2   & 0.115&0.11 &0.428\\
$\Upsilon_b$ & 3.59	& $0.6^{3}$ 	& 0.8  & 5 & 0.36   &1.2 &0.115 &0.016 &0.638\\
\hline
\end{tabular}
\end{table*}

\begin{figure}[t]
\hspace*{-0.05\columnwidth}\includegraphics[width=1.1\columnwidth]{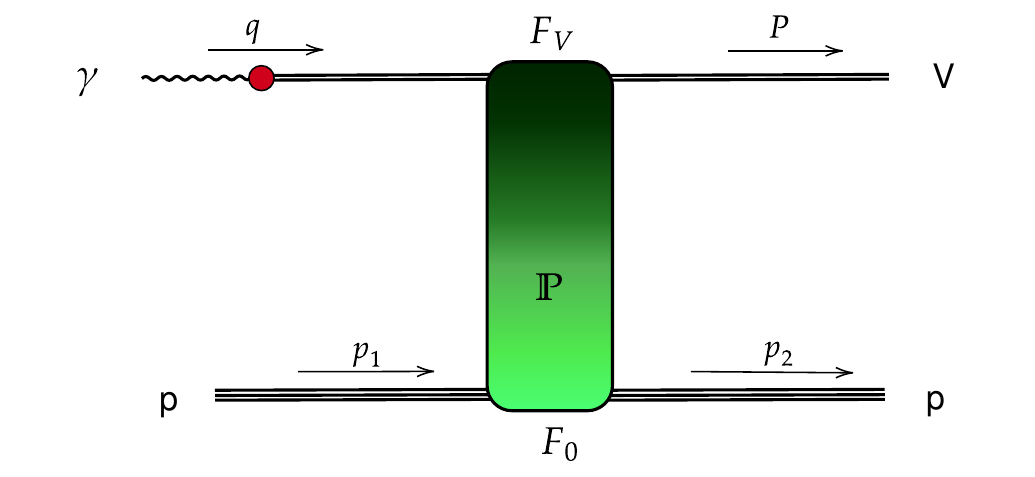}
\caption{\label{Pomam}
$\mathbb{P}-$am reaction model for $\gamma + p \to V + p$.
Compared with our dynamical model, $\mathbb{P}-$dyn, sketched in Fig.\,\ref{FigPomMechanism}, $\mathbb{P}-$am replaces the $\gamma\to {\mathpzc q} \bar {\mathpzc q} + \mathbb P \to V$ transition matrix element by a momentum-independent $\gamma\to V$ transition coupling (akin to vector meson dominance) along with a form factor, $F_V$, that expresses a Pomeron--vector-meson form factor.
}
\end{figure}

\subsection{VMD-like treatment of photon to meson transition}
\label{sec:23}
In the following, we will compare our predictions with those obtained using a simplified framework that replaces the calculated $\gamma\to {\mathpzc q} \bar {\mathpzc q} + \mathbb P \to V$ transition matrix element by a phenomenologically constrained model.
Sketched in Fig.\,\ref{Pomam} and identified as the $\mathbb P\,-$am model in Ref.\,\cite{Tang:2024pky}, this approach is obtained by beginning with the Ref.\,\cite{Sakinah:2024cza} model;
setting $v_{cN}= 0$, so the quark-loop integration, based on a constituent quark model wavefunction, and $V-N$ final-state interactions (FSIs) do not contribute to the amplitude;
and updating the Pomeron trajectories and Pomeron-quark couplings.

The associated matrix element for diffractive vector meson photoproduction is
\cite[Eqs.\,(41-43)]{Sakinah:2024cza}:
\begin{align}
{\cal I}_\mu(W,t)\propto& \,\bar{u}(p_2)\gamma_\alpha u(p_1) \epsilon^\lambda_\nu(P) \Gamma_{\alpha,\mu\nu} G_{\mathbb P}(w^2,t)   \notag\\
&\times [\beta_{\mathpzc q} F_{V}(t)][\beta_{l} F_{0}(t)]\,,
\end{align}
with the tensor structure of the $\mathbb P$-$V$ vertex being a minimal \emph{Ansatz} that ensures physical constraints \cite[Sec.\,III.A]{Titov:1998bw}, \emph{viz}.
\begin{equation}
\Gamma_{\alpha,\mu\nu}=q_\alpha \delta_{\mu\nu} - q_\mu \delta_{\alpha\nu}\,.
\end{equation}
The constants $\beta_{{\mathpzc q},\ell}$ fix the strengths of $\mathbb P$ interactions with the quarks in the vector mesons and the proton, respectively.
Drawn from Ref.\,\cite{Sakinah:2024cza}, the phenomenological form factor
\begin{equation}
 F_{V}(t)=\frac{1}{m_{V}^{2}-t}\left(\frac{2\mu_{0}^{2}}{2\mu_{0}^{2}+m_{V}^{2}-t}\right),
 \label{PomamFV}
\end{equation}
$\mu_{0}=1.1\,$GeV, replaces the momentum dependence of the $\gamma\to {\mathpzc q} \bar {\mathpzc q} + \mathbb P \to V$ loop and vastly simplifies its tensor structure.

\subsection{Pomeron trajectories and couplings}
\label{sec:24}
In the subsequent calculations, we adopt Pomeron trajectories that are either well-established or reliably inferred, enabling accurate descriptions of existing experimental data \cite{Pichowsky:1996tn, ZEUS:2002wfj, Capua:2012sd}.
A common trajectory is used for light mesons and another for the heavy systems; see Table~\ref{tab:1}.
The coupling $\beta_{u}$ is determined via a least-squares fit to the differential cross section for $\rho^{0}$ photoproduction \cite[ZEUS1995]{ZEUS:1995bfs}.
(The $\beta_u$ value used herein is 25\% larger than that in Ref.\,\cite{Tang:2024pky} as a consequence of correcting a coding error in that analysis.)
The couplings $\beta_{s}$ and $\beta_{b}$ are obtained through fits to high-energy total cross-section data on $\phi$ \cite{ZEUS:1996esk} and $\Upsilon$ production \cite{ZEUS:1998cdr, H1:2000kis, ZEUS:2009asc, LHCb:2015wlx, CMS:2018bbk};
and the $\beta_{c}$ values are fixed, as in Ref.\,\cite{Tang:2024pky}, by requiring a best fit to the $W$-dependence of the $|t|\approx |t|_{\rm min}$ $J/\psi$ photoproduction differential cross section in Ref.\,\cite[ZEUS\,2002]{ZEUS:2002wfj}.

\section{Photon to vector meson transition amplitude in CSMs}
\label{sec:3}
The photoproduction reaction mechanism sketched in Fig.\,\ref{FigPomMechanism} involves the
dressed quark propagator,
dressed photon-quark vertex;
and vector-meson Bethe-Salpeter amplitude.
Such quantities have long been the focus of CSM studies and many relevant reviews are available; see, \emph{e.g}., Refs.\,\cite{Roberts:2012sv, Burkert:2017djo, Bashir:2012fs, Eichmann:2016yit, Roberts:2021nhw, Binosi:2022djx, Ding:2022ows, Ferreira:2023fva, Raya:2024ejx, Achenbach:2025kfx}.
Hence, today, these quantities are readily computed, so it is possible to arrive at parameter-free predictions for the $\gamma\to {\mathpzc q} \bar {\mathpzc q} + \mathbb P \to V$ transition matrix elements, $t_{\mu \alpha \nu}(q,P)$ in Eq.\,\eqref{EqtqP}.  The key to such calculations is the quark+antiquark scattering kernel. For this, the leading-order (rainbow-ladder, RL) truncation \cite{Munczek:1994zz, Bender:1996bb} is obtained by writing \cite{Maris:1997tm}:
{\allowdisplaybreaks
\label{EqRLInteraction}
\begin{align}
\label{KDinteraction}
\mathscr{K}_{tu}^{rs}(k) & =
\tilde{\mathpzc G}(y)
[i\gamma_\mu\frac{\lambda^{a}}{2} ]_{ts} [i\gamma_\nu\frac{\lambda^{a}}{2} ]_{ru} T_{\mu\nu}(k)\,,
%
\end{align}
$k^2 T_{\mu\nu}(k) = k^2 \delta_{\mu\nu} - k_\mu k_\nu$,  $y=k^2$.  The tensor structure corresponds to Landau gauge, used because it is both a fixed point of the renormalisation group and that gauge for which corrections to RL truncation are least noticeable \cite{Bashir:2009fv}.
In Eq.\,\eqref{EqRLInteraction}, $r,s,t,u$ represent colour, spinor, flavour matrix indices (as necessary).
}
A realistic form of $\tilde{\mathpzc G}(y) $ is explained in Refs.\,\cite{Qin:2011dd, Binosi:2014aea}:
\begin{align}
\label{defcalG}
 \tilde{\mathpzc G}(y) & =
 \frac{8\pi^2}{\omega^4} D e^{-y/\omega^2} + \frac{8\pi^2 \gamma_m \mathcal{F}(y)}{\ln\big[ \tau+(1+y/\Lambda_{\rm QCD}^2)^2 \big]}\,,
\end{align}
${\cal F}(y) = \{1 - \exp(-y/\Lambda_{\mathpzc I}^2)\}/y$, $\Lambda_{\mathpzc I}=1\,$GeV;
$\tau={\rm e}^2-1$;
$\gamma_m=12/(33-2 N_f)$, with $N_f$ and $\Lambda_{\rm QCD}^{N_f}$ given in Table~\ref{tab:1}.

At this point, with the quark current masses and parameter values in Table~\ref{tab:1}\,left-block, the necessary gap and Bethe-Salpeter equations can be solved using standard algorithms \cite{Maris:1997tm, Maris:2005tt, Krassnigg:2009gd} to obtain each element in Eq.\,\eqref{EqtqP} and therewith $t_{\mu \alpha \nu}^V(q,P)$ itself.
In solving all relevant Schwinger function equations, we use a mass-independent (chiral-limit) momen\-tum-subtraction renormalisation scheme \cite{Chang:2008ec}, with re\-normalisation scale $\zeta=19\,$GeV.  At this scale, the quark wave function renormalisation constants are practically unity.

Applications to numerous systems and reactions have established \cite{Ding:2022ows} that interactions in the class typified by Eqs.\,\eqref{EqRLInteraction}, \eqref{defcalG} can unify the properties of many systems.
Notably, when $\omega D=:\varsigma^3$ is held fixed, results for observable quantities remain practically unchanged under $\omega \to (1\pm 0.2)\omega$ \cite{Qin:2020rad}.
Thus, the interaction in Eq.\,\eqref{defcalG} is determined by just one parameter: $\varsigma^3:=\omega D$.

\begin{table}[t]
\centering
\caption{Pion and vector meson masses and leptonic decay constants computed herein, using CSMs, compared with experimental values \cite[PDG]{ParticleDataGroup:2024cfk}, quoted at a sensible level of precision.
Regarding vector mesons, the mean absolute relative difference between prediction and experiment is $5(3)$\%.}
\label{TabStatic}
\begin{tabular}{>{\centering\arraybackslash}p{0.07\textwidth}|>{\centering\arraybackslash}p{0.07\textwidth}>{\centering\arraybackslash}p{0.07\textwidth}|>{\centering\arraybackslash}p{0.07\textwidth}>{\centering\arraybackslash}p{0.07\textwidth}}
\hline
Meson & \multicolumn{2}{|c}{Mass [GeV] } & \multicolumn{2}{|c}{Decay constant [GeV]}  \\
\noalign{\smallskip}\hline\noalign{\smallskip}
 & Herein & Expt. & Herein & Expt. \\
\noalign{\smallskip}\hline\noalign{\smallskip}
$\pi$ & 0.138 & 0.138 & 0.092 & 0.092 \\
$\rho^{0}$ & 0.741 & 0.775 & 0.149 & 0.153 \\
$\phi$ & 1.086 & 1.019 & 0.183 & 0.168 \\
$J/\psi$ & 3.123 & 3.097 & 0.278 & 0.294 \\
$\Upsilon(1s)$ & 9.52$\ $ & 9.46$\ $ & 0.558 & 0.51$\ $ \\
\hline
\end{tabular}
\end{table}

\begin{figure}[t]
\leftline{\hspace*{0.5em}{\large{\textsf{A}}}}
\vspace*{-2ex}
\centerline{\includegraphics[width=0.45\textwidth]{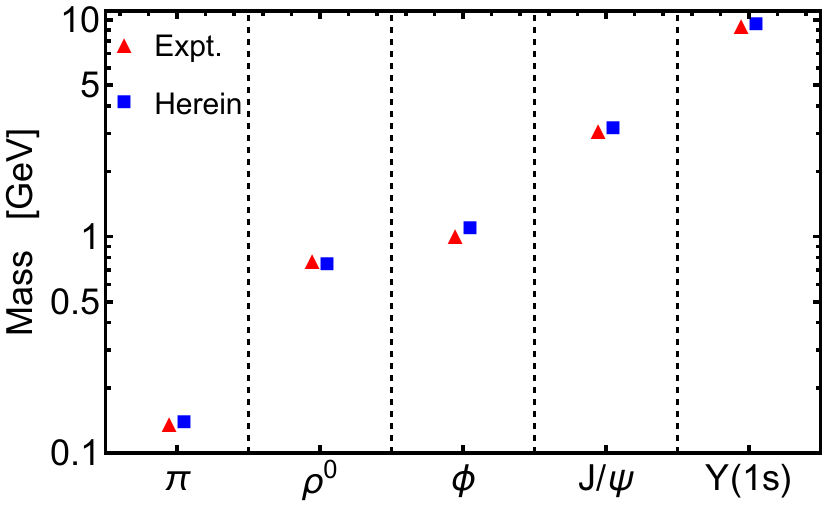}}

\vspace*{1ex}

\leftline{\hspace*{0.5em}{\large{\textsf{B}}}}
\vspace*{-1.5ex}
\centerline{\includegraphics[width=0.45\textwidth]{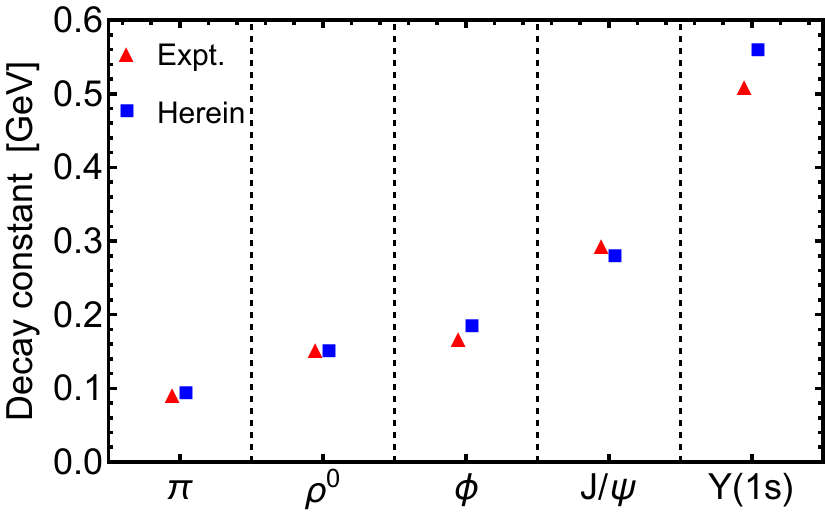}}

\caption{Pion and vector meson properties listed in Table~\ref{TabStatic}, computed using CSMs with interaction parameters in Table~\ref{tab:1} \cite{Qin:2018dqp, Xu:2019ilh, Yao:2021pyf}: ({\sf A}) Masses and ({\sf B}) leptonic decay constants.
Blue squares indicate CSM results; red triangles indicate experimental values at a sensible level of precision \cite{ParticleDataGroup:2024cfk}.
\label{fig:2}}
\end{figure}

In the light-quark sector, $f=u, d, s$, $\varsigma_f  =0.8\,{\rm GeV}$ delivers a good description of a range of pseudoscalar and vector meson static properties; see Table~\ref{TabStatic} and Fig.\,\ref{fig:2}.
These results were obtained as explained, \emph{e.g}., in Refs.\,\cite{Chen:2018rwz, Ding:2018xwy, Xu:2019ilh}.  As noted above, the procedure involves solving sets of coupled gap and Bethe-Salpeter equations.

For heavier quarks, it is known that corrections to RL truncation are less important -- see, \emph{e.g}.,  Ref.\,\cite{Bhagwat:2004hn} for illustration, so the interaction parameter should be closer to that used in more sophisticated kernels.  This is discussed further in, \emph{e.g}., Ref.\,\cite[Sect.\,IIB]{Xu:2019ilh}, and explains the $c, b$ quark values listed in Table~\ref{tab:1}, which deliver the results listed in Table~\ref{TabStatic} and drawn in Fig.\,\ref{fig:2}.

\section{Near-threshold slope parameter / meson-proton scattering lengths}
\label{sec:4}
Diffractive exclusive vector meson photoproduction \linebreak near threshold provides insights into the low-energy meson-proton interaction, which is often characterised by a scattering length.
Despite its failings \cite{Xu:2021mju}, especially in applications to heavier vector mesons, \linebreak many phenomenological studies continue to use vector meson dominance as a phenomenological tool.
Within that framework, the photoproduction cross section\linebreak $\sigma_{\gamma p\to Vp}$ is simply related to the meson-proton elastic scattering cross section, $\sigma_{Vp\to Vp}$, such that, near threshold, the differential cross section can be expressed as follows \cite{Lee:2022ymp}:
{\allowdisplaybreaks
\begin{eqnarray}
     \frac{d\sigma^{\gamma p \to Vp}}{d\Omega}\bigg{|}_{\text{th}}&=&\frac{|\vec{P}|}{k_W}\frac{1}{64\pi}|T^{\gamma p\to Vp}|^{2},\notag \\
    &=&\frac{|\vec{P}|}{k_W}\frac{\pi \alpha}{g_{V}^{2}}\frac{d\sigma^{Vp\to Vp}}{d\Omega}\bigg{|}_{\text{th}}\notag \\
    &=&\frac{|\vec{P}|}{k_W}\frac{\pi \alpha}{g_{V}^{2}}|\alpha_{Vp}|^{2}, \label{eq:19}
\end{eqnarray}
where $\alpha$ is the fine-structure constant,
$\vec{P}$ is the final momentum of the vector meson
-- Eq.\eqref{Eqmomenta},
and $\alpha_{Vp}$ is the model meson-proton scattering length.
The VMD ingredient is the coupling $g_{V}$, which is defined by the vector meson's electromagnetic decay width $\Gamma_{V\to e^{+}e^{-}}$:
\begin{equation}
    g_{V}^{2}=\frac{\pi\alpha^{2}m_{V}}{3\Gamma_{V\to\text{e}^{+}\text{e}^{-}}}.
\end{equation}
}

Suppose now that $V$-meson production near threshold is purely an $S$-wave process, then $\sigma_{\rm th} = 4\pi d\sigma/d\Omega|_{\rm th}$.
Then if one has precise total cross section data in hand, very close to threshold, it is possible to fit that data using a polynomial in $|\vec{P}|$, \emph{viz}.\ write \cite{Strakovsky:2019bev}
\begin{equation}
\sigma_{\rm th} = {\mathpzc a}_1 |\vec{P}| + {\mathpzc a}_3 |\vec{P}|^3 + {\mathpzc a}_5 |\vec{P}|^5 ,
\label{sigmafit}
\end{equation}
whereafter one has
\begin{equation}
|\alpha_{Vp}|^2  = {\mathpzc a}_1 \frac{ g_V^2 k_W }{4\pi^2 \alpha}\,. \label{INew}
\end{equation}

Our prediction for the differential cross section is given in Eq.\,\eqref{eq:7}.
That is the formula we use to evaluate all results.
However, by equating Eq.\,\eqref{eq:7} with Eq.\,\eqref{eq:19} one can arrive at model-specific VMD values for the meson-proton scattering length:
\begin{equation}
|\alpha_{Vp}|^{2}  =
\frac{1}{16\pi^2} \frac{m_p}{W}
\frac{g_V^2}{\pi\alpha}
\, \tfrac{1}{2}\!\!\!\sum_{\text{proton spin}} \!\!\!
\left.   (|{\cal I}_1|^{2}+|{\cal I}_2|^{2})\right|_{\rm th}
\,.\label{avmd1}
\end{equation}
These can be used as single parameters with which to evaluate the veracity of existing such model dependent inferences of meson-proton scattering lengths \cite{Strakovsky:2019bev, Strakovsky:2020uqs, Strakovsky:2021vyk, Strakovsky:2025ews}.
Furthermore, obtained in this way, one need not interpret $|\alpha_{Vp}|$ as the $V-p$ scattering length. Instead, it can be viewed as a near-threshold slope parameter.  Then, the value has objective utility as a comparison between theory and experiment.
We stress that one must use data very close to threshold if the fitted value of $\mathpzc a_1$ in Eq.\,\eqref{sigmafit} is to deliver a reliable empirical result for the slope parameter in Eq.\,\eqref{avmd1}.


\section{Results: cross sections}
\label{sec:5}
%
With the quark current-masses and interaction parameters, Table~\ref{tab:1} -- and the propagators and Bethe-Salpeter amplitudes they produce -- validated in comparisons with experimental values for meson masses and decay constants, Table~\ref{TabStatic} and Fig.\,\ref{fig:2}, we use these elements to calculate the matrix elements defined by Eqs.\,\eqref{eq:1}, \eqref{EqtqP}.

Accounting for current conservation and the fact that $P_\nu \Gamma_\nu^V = 0$, then, in all cases, $t_{\mu\alpha\nu}^V$ involves $9$ independent scalar functions which modulate the strength of the allowed tensor structures:
\begin{equation}
t_{\mu\nu\alpha}^V(q,P) = \sum_{i=1}^{9}F_{i}^V(q,P)\tau^{i}_{\mu\nu\alpha}(q,P).
\label{Eqtmunualpha}
\end{equation}
The nine tensors are listed explicitly in \ref{App1}.

\begin{figure}[t]
  \centering
  \includegraphics[width=0.475\textwidth]{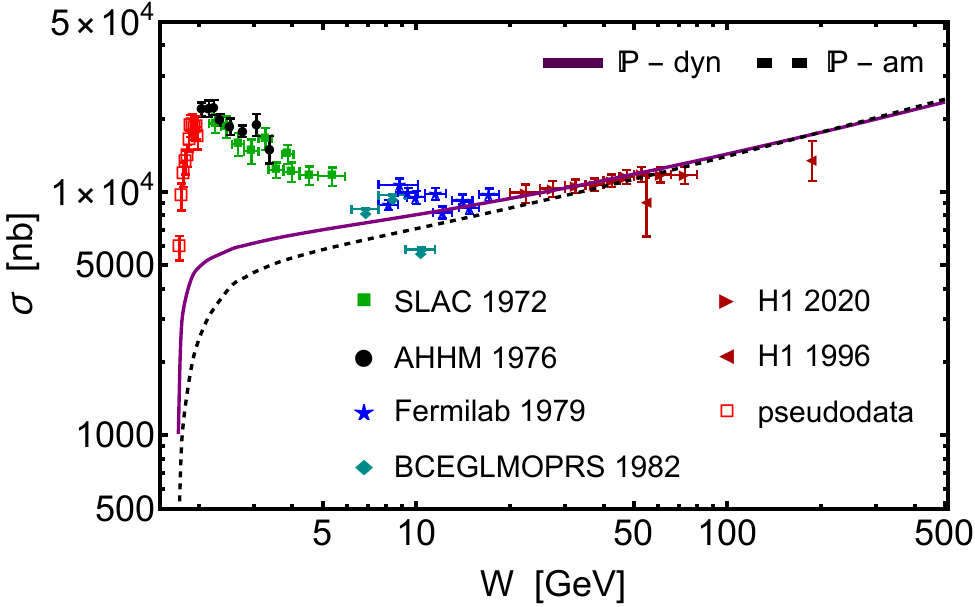}
  \caption{$W$-dependence of total cross section for $\rho^{0}$ photoproduction, $W_{\rm th}^{\rho} = 1.714\,$GeV.
  Solid purple curve -- $\mathbb{P}-$dyn;
  dashed black curve -- $\mathbb{P}-$am.
  Data sources: black circles -- \cite[AHHM]{Aachen-Hamburg-Heidelberg-Munich:1975jed};
   cyan diamonds (fixed target) -- \cite[BCEGLMOPRS]{Bonn-CERN-EcolePoly-Glasgow-Lancaster-Manchester-Orsay-Paris-Rutherford-Sheffield:1982qiv};
   green squares \cite[SLAC]{Park:1971ts, Ballam:1971wq};
   blue stars \cite[FNAL]{Egloff:1979mg};
   red left and right triangles \cite[H1]{H1:1996prv, H1:2020lzc};
   and red open squares -- pseudodata inferred \cite{Strakovsky:2025ews}
   from differential $\pi^+\pi^-p$ photoproduction cross sections measured with CLAS at JLab and analysed within the JM model \cite{Mokeev:2015lda} (see text).
  \label{fig:4}}
\end{figure}

Each scalar function in Eq.\,\eqref{Eqtmunualpha} is readily extracted using sensibly chosen projection operators.
In practice, we find that the scalar function associated with the tensor structure
\begin{equation}
\tau^{1}_{\mu\nu\alpha}(q,P)=q_{\alpha}(\delta_{\mu\nu}-P_{\mu}P_{\nu}/P^{2})
\end{equation}
gives the dominant contribution to the photoproduction cross sections in all cases: for heavy mesons, $J/\psi$, $\Upsilon$, all other tensors can be neglected; and for light mesons, $\rho$, $\phi$, it provides roughly 85\% of the cross section, with the tensor
\begin{equation}
\tau^{2}_{\mu\nu\alpha}(q,P) = P_{\alpha}(\delta_{\mu\nu}-P_{\mu}P_{\nu}/P^{2})
\end{equation}
contributing the remaining 15\% -- alone (5\%) and via constructive interference with $\tau^{1}$ (10\%).
Notwithstanding these observations, all tensor structures are employed to produce the results reported herein.

For each vector meson, we calculate both the differential and total cross sections for exclusive photoproduction.
The differential cross section depends on $W$, the centre-of-mass energy, and $t$, the square-momentum transfer.
When relevant data are available, we examine both the $W$- and $t$-dependence.
Often, the $W$ dependence is analysed at $|t|=|t|_{\text{min}}$, Eq.\,\eqref{Eqtmin}, which corresponds to forward scattering, \emph{i.e}., the smallest possible value for which the vector meson can be produced leaving the proton in tact.
The cross sections are largest on this domain and very sensitive to reaction details.
For instance, within our approach, they are much affected by the quark-loop dynamics expressed by Fig.\,\ref{FigPomMechanism}, as shown for the $J/\psi$ in Ref.\,\cite{Tang:2024pky} and we will see again herein for other vector mesons.
On the other hand, the $t$-dependence of the differential cross sections are usually mapped at fixed $W$.  Near threshold, they, too, are sensitive to the vector meson production mechanism.

It is worth noting here that in fixed-target experiments, the photon energy is often used as the primary variable, \emph{viz}.\ $E_\gamma(W) = k_W$ in Eq.\,\eqref{eqkW}.

\begin{figure}[t]
  \centering
   \includegraphics[width=0.475\textwidth]{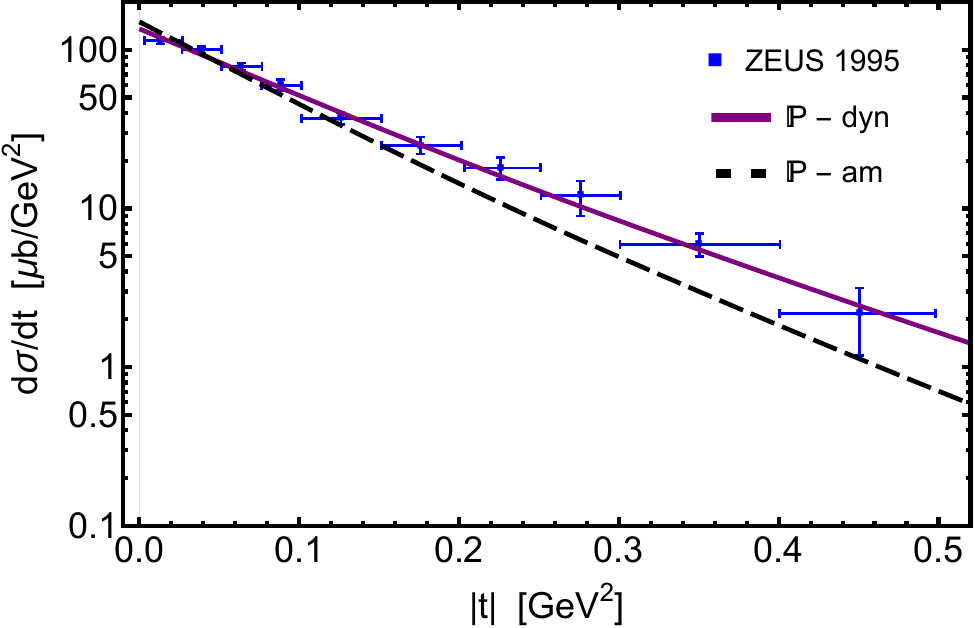}
  \caption{Differential cross section for $\rho^{0}$ photoproduction compared with data \cite[ZEUS]{ZEUS:1995bfs} (blue squares) in the range $60<W<$ $80$ GeV.
  Solid purple curve -- $\mathbb{P}-$dyn; dashed black curve: $\mathbb{P}-$am.
  Both are computed at the average energy $\text{W}=70$ GeV.  Pomeron trajectory and quark–Pomeron couplings used are given in Table~\ref{tab:1}\,--\,Row~1. \label{fig:3}}
\end{figure}

\subsection{$\rho$ photoproduction}
\label{SSrho}
%
Calculated using the Pomeron trajectory and quark–Po\-meron couplings given in Table~\ref{tab:1}\,--\,Row~1, our predictions for $\gamma + p \to \rho^{0} + p$ photoproduction are presented in Figs.\,\ref{fig:4} and \ref{fig:3}.

In Fig.\,\ref{fig:4}, we draw the total cross section on $W\in [W_{\rm th}^{\rho},500\,{\rm GeV}]$ along with data from a variety of sources
\cite[AHHM]{Aachen-Hamburg-Heidelberg-Munich:1975jed}, \cite[BCEGLMOPRS]{Bonn-CERN-EcolePoly-Glasgow-Lancaster-Manchester-Orsay-Paris-Rutherford-Sheffield:1982qiv}, \cite[SLAC]{Park:1971ts, Ballam:1971wq},
\cite[FNAL]{Egloff:1979mg},
\cite[H1]{H1:1996prv, H1:2020lzc},
plus pseudodata obtained from differential $\pi^+\pi^-p$ photoproduction cross sections measured with the CLAS detector at JLab, analysed within the JLab-Moscow State University (JM) model \cite{Mokeev:2015lda},
and integrated over the differential solid angle.
Both $\mathbb{P}-$dyn and $\mathbb{P}-$ am reproduce the data on $W\gtrsim 10\,$GeV, which may therefore be identified as the domain of Pomeron dominance in this case.
Furthermore, the $\mathbb P-$dyn model is satisfactory on $W\gtrsim 5\,{\rm GeV}$, highlighting the importance of a sound description of the quark loop in Fig.\,\ref{FigPomMechanism}.
However, both models are equally poor on $W\in [W_{\rm th}^{\rho},5\,{\rm GeV}]$.
This is because the proton and $\rho$ meson share valence quarks; hence, many additional (nondiffractive) ex\-change contributions are possible in the photoproduction reaction.
For instance, a leading-order calculation using a phenomenological meson-ex\-change model \cite{Sato:1996gk} is able to reproduce the low-$W$ data \cite[Fig.\,15]{Pichowsky:1996tn}.
Such contributions decay rapidly with $W$, so that the diffractive $\mathbb{P}-$dyn reaction model becomes valid on $W\gtrsim 5\,$GeV.

\begin{figure}[t]
\centering
\includegraphics[width=0.475\textwidth]{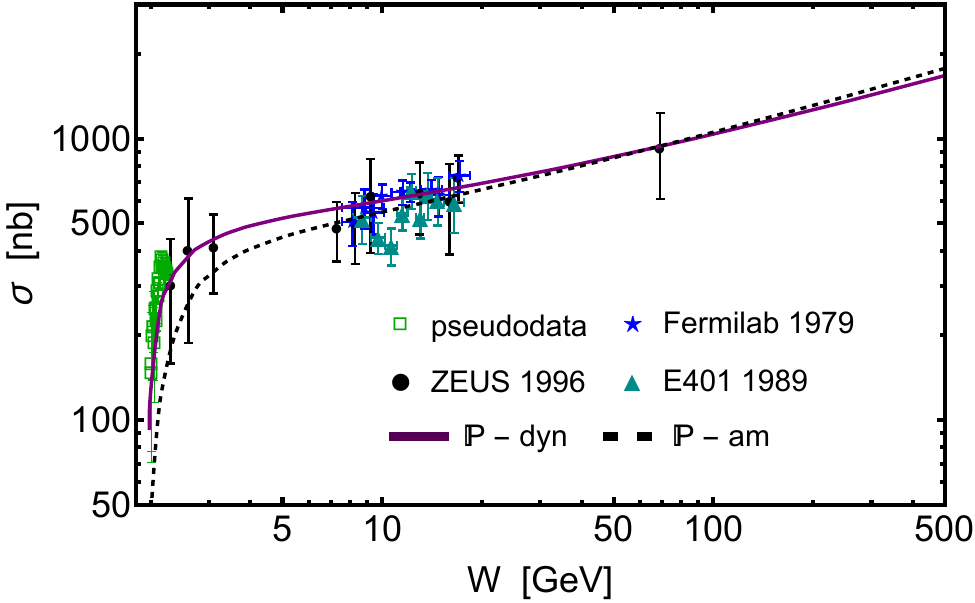}
\caption{$W$-dependence of total cross section for $\phi$ photoproduction, $W_{\rm th}^\phi = 1.958\,$GeV.
Solid purple curve -- $\mathbb{P}-$dyn; dashed black curve -- $\mathbb{P}-$am.
Data: blue stars -- \cite[FNAL]{Egloff:1979mg};
cyan triangles -- \cite[E401]{Busenitz:1989gq};
black circles -- \cite[ZEUS]{ZEUS:1996esk};
and green open squares ``pseudodata'' \cite{Strakovsky:2020uqs}, obtained by interpolating CLAS differential cross sections \cite[CLAS]{Dey:2014tfa} as described in the text.
  \label{fig:9}}
\end{figure}

Figure~\ref{fig:3} depicts the differential cross section.
It compares $\mathbb{P}-$dyn and $\mathbb{P}-$am predictions with data from Ref.\,\cite[ZEUS]{ZEUS:1995bfs} in the range $60 <\text{W}<$ $80$ GeV.
Our results are evaluated at the average experimental energy, $W = 70$ GeV.
Recalling that the Pomeron trajectory parameters in Table~\ref{tab:1}\,--\,Row 1 were taken from Ref.\,\cite{Pichowsky:1996tn} and only the quark-Pomeron couplings were tuned to the data in Fig.\,\ref{fig:4} -- see Sec.\,\ref{sec:24} -- then it becomes clear that the quark-loop dynamics in Fig.\,\ref{FigPomMechanism} has an impact on the differential cross section even far above threshold, $W\gg W_{\rm th}^\rho$.

 Regarding $\omega$-meson photoproduction, owing to its comparable structure, similar features and explanations pertain to the cross sections \cite{Yu:2017vvp}.

\begin{figure}[t]
  \centering
  \includegraphics[width=0.5\textwidth]{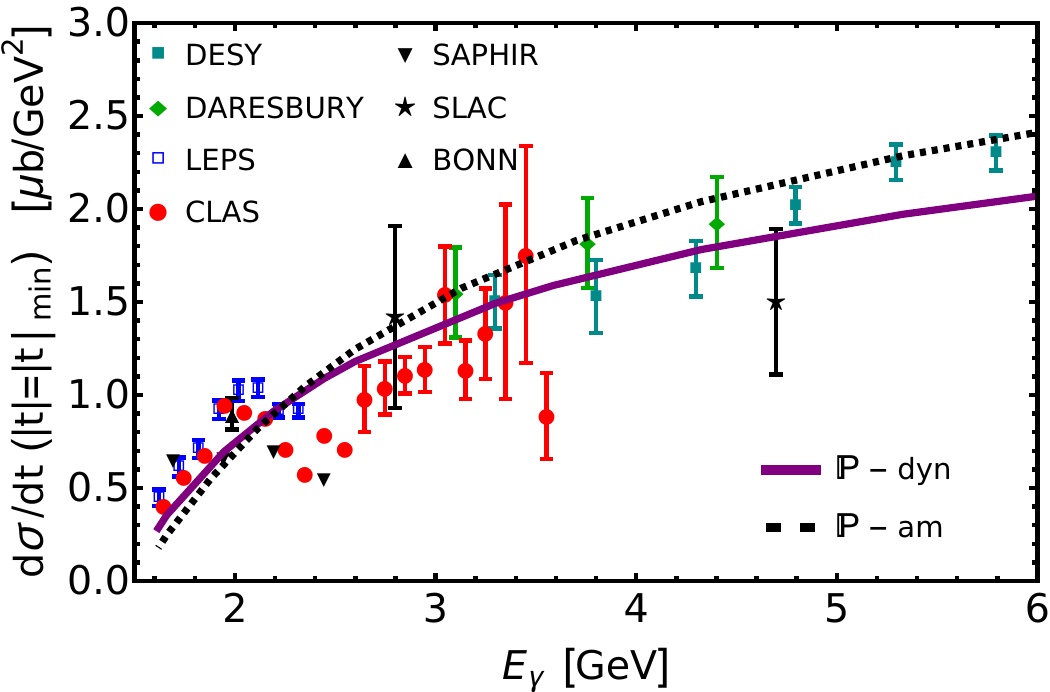}
  \caption{$E_\gamma$ dependence of differential cross section for $\phi$ photoproduction, evaluated at $|t| = |t|_{\text{min}}$,  Eq.\,\eqref{Eqtmin}, corresponding to forward-angle scattering.
  Solid purple curve -- $\mathbb{P}-$dyn; dashed black curve -- $\mathbb{P}-$am.
  Data: black stars -- \cite[SLAC]{Ballam:1972eq};
  black up triangles -- \cite[Bonn]{Besch:1974rp};
  cyan squares -- \cite[DESY]{Behrend:1978ik};
  green -- \cite[Daresbury]{Barber:1981fj};
  black down triangles -- \cite[SAPHIR]{Barth:2003bq};
  blue open squares -- \cite[LEPS]{LEPS:2005hax};
  red circles -- \cite[CLAS]{CLAS:2013jlg}.
  \label{fig:5}}
\end{figure}

\subsection{$\phi$ photoproduction}
\label{SSphi}

\begin{figure*}[t]
  \centering
\includegraphics[width=0.24\textwidth]{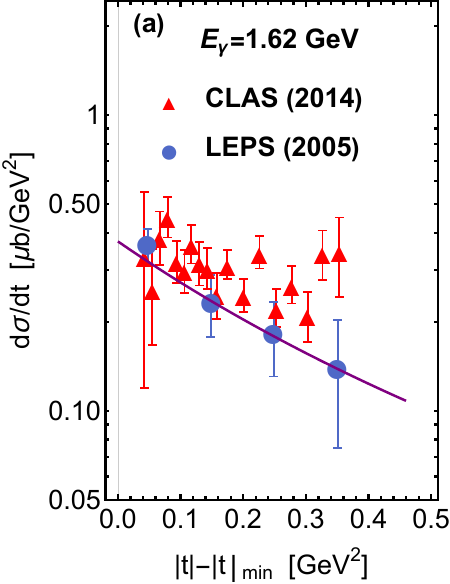}
\includegraphics[width=0.24\textwidth]{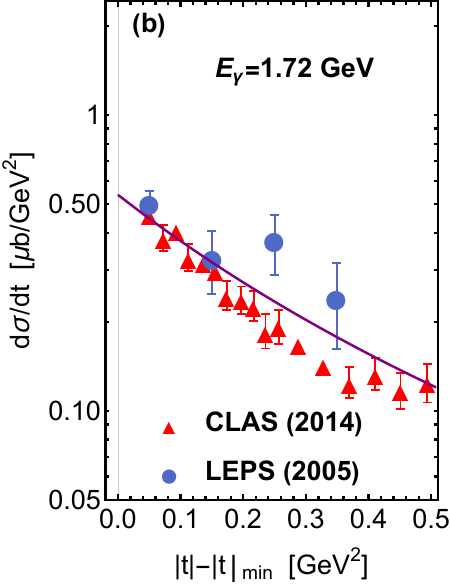}
\includegraphics[width=0.24\textwidth]{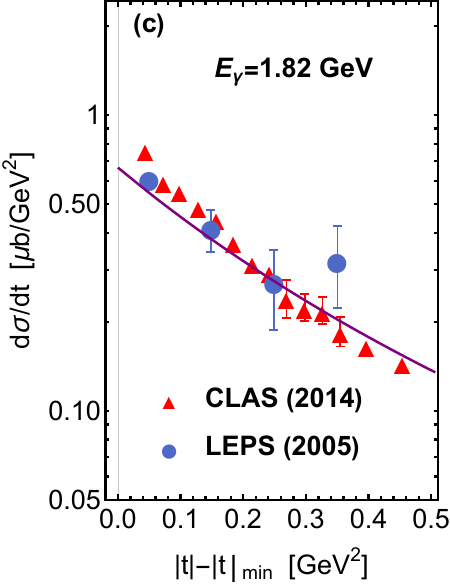}
\includegraphics[width=0.24\textwidth]{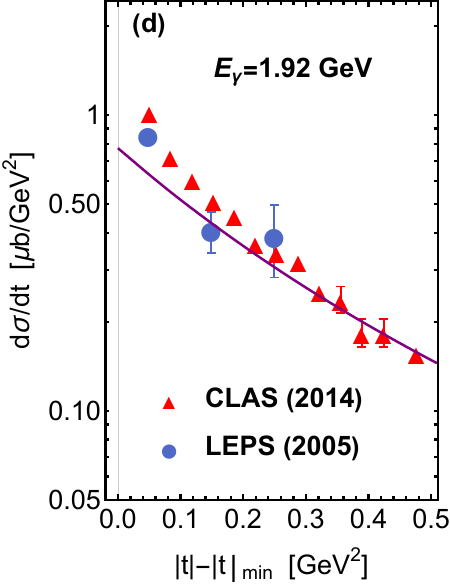}\\
\includegraphics[width=0.24\textwidth]{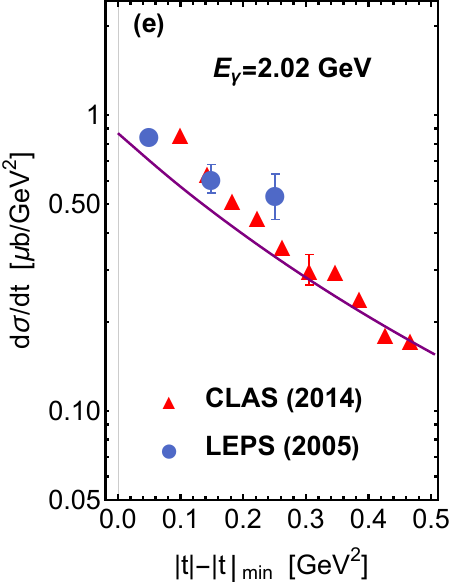}
\includegraphics[width=0.24\textwidth]{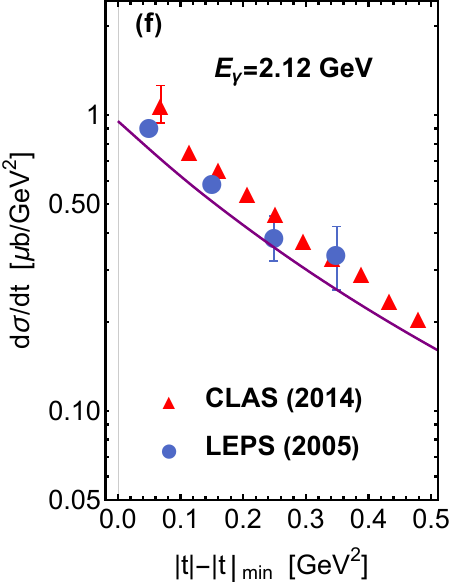}
\includegraphics[width=0.24\textwidth]{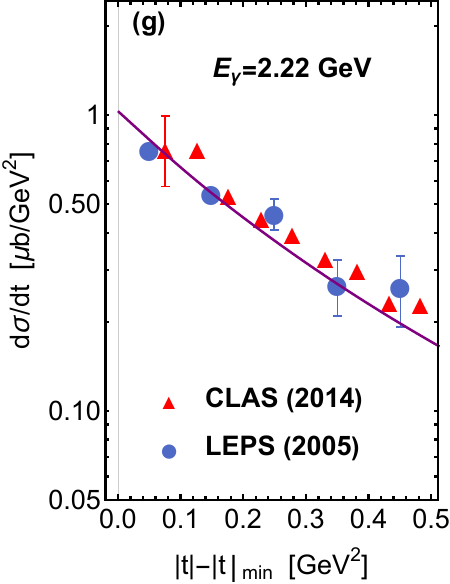}
\includegraphics[width=0.24\textwidth]{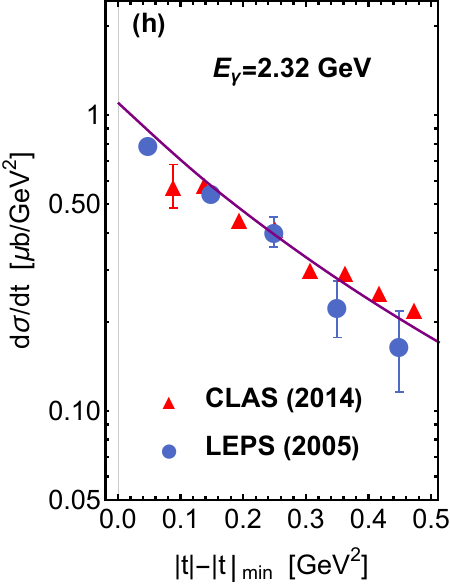}
\caption{Differential cross sections for $\phi$ photoproduction as a function of $|t|-|t|_{\text{min}}$ at low photon energies, $E_\gamma$.
Solid purple curve  -- $\mathbb{P}-$dyn.
Data: red up-triangles -- \cite[CLAS\,2014]{Dey:2014tfa};
blue circles -- \cite[LEPS]{LEPS:2005hax}.
  \label{fig:8}}
\end{figure*}

Using the Pomeron trajectory and quark-Pomeron couplings given in Table~\ref{tab:1}\,--\,Row~2, we obtain the predictions for $\gamma + p \to \phi + p$ photoproduction cross sections presented in Figs.\,\ref{fig:9}\,--\,\ref{fig:8}.
An important difference between this reaction and $\rho^0$ photoproduction is that the $\phi$-meson contains no valence quarks in common with the proton; hence, fewer nondiffractive processes are possible.

The total cross section, $\sigma_{\gamma p \to \phi p}$, is drawn in Fig.\,\ref{fig:9}.
Comparing the $\mathbb{P}-$dyn and $\mathbb{P}-$am predictions with data, it is evident that the dynamical quark loop associated with Fig.\,\ref{FigPomMechanism} is crucial in order to obtain agreement with data.
This is especially true near threshold.
Notably, very near threshold, direct measurements of $\sigma_{\gamma p \to \phi p}$ are unavailable.
To fill that gap, Ref.\,\cite{Strakovsky:2020uqs} generated pseudodata by interpolating differential cross sections measured at JLab \cite[CLAS]{Dey:2014tfa} and integrating over the differential solid angle.
The $\mathbb{P}-$dyn prediction aligns well with this pseudodata, too; although it misses the ``bump''
that inspection reveals in the pseudodata \cite[Fig.\,1]{Strakovsky:2020uqs}.
We note that this bump may be merely an artefact of the pseudodata construction procedure.

The $E_\gamma$ dependence of the forward-angle $\phi$ photoproduction differential cross section is drawn in Fig.\,\ref{fig:5}.
Compared with data, the $\mathbb{P}-$dyn reaction model delivers a better global description than $\mathbb{P}-$am; hence, our dynamical treatment of the $\gamma\to {\mathpzc q}\bar {\mathpzc q} +\mathbb{P}\to \phi$ transition is once more seen to be important.

It should nevertheless be remarked that some structure seems apparent in Fig.\,\ref{fig:5} on a domain $E_\gamma \simeq 2.5\,$GeV.
This is reproduced by neither our reaction model nor an earlier analysis \cite{Titov:2003bk} that also included $s$-channel resonance contributions with subsequent $\phi$-meson radiation.
Since hadron interactions in these channels are poorly known, it might be that the structure owes, at least in part, to singularities associated, \emph{e.g}., with meson + baryon FSIs \cite{Ryu:2012tw}.
All that can now be safely concluded is that diffractive mechanisms do not reproduce $E_\gamma \simeq 2.5\,$GeV structure, and more data are required before it may properly be understood.

\begin{figure}[t]
\leftline{\hspace*{0.5em}{\large{\textsf{A}}}}
\vspace*{-2ex}
\centerline{\includegraphics[width=0.475\textwidth]{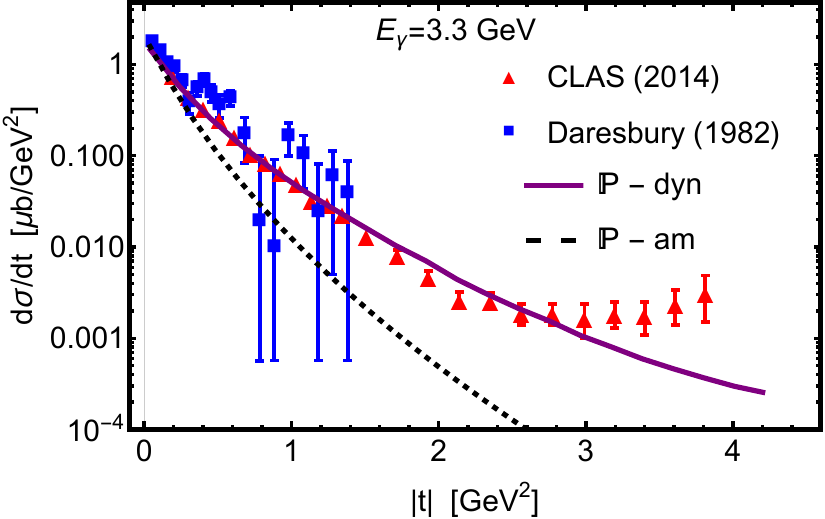}}

\vspace*{1ex}

\leftline{\hspace*{0.5em}{\large{\textsf{B}}}}
\vspace*{-1.5ex}
\centerline{\includegraphics[width=0.475\textwidth]{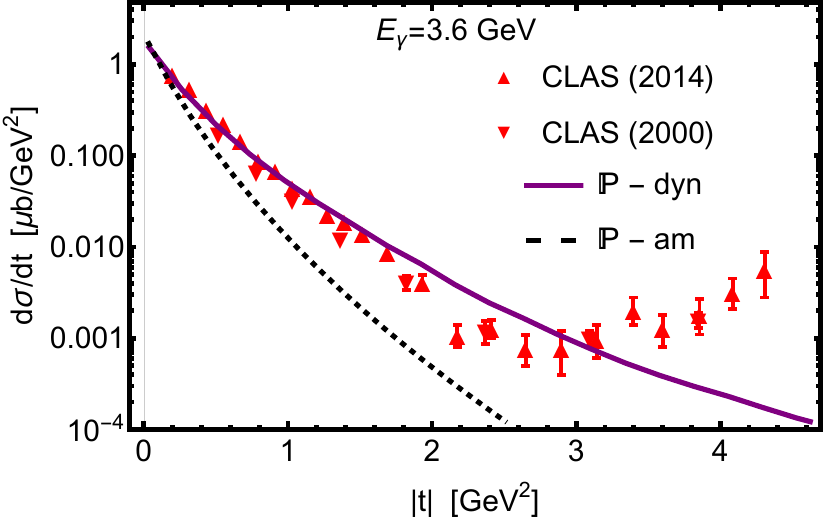}}

  \caption{$|t|$ dependence of differential cross section for $\phi$ photoproduction at photon energies: $E_{\gamma} = 3.3\,$GeV ($W \approx 2.66$ GeV) and $E_{\gamma} = 3.6\,$GeV ($W \approx 2.76$ GeV).
  Solid purple curve  -- $\mathbb{P}-$dyn;
  dashed black curve -- $\mathbb{P}$ - am.
  Data:
  red up-triangles -- \cite[CLAS\,2014]{Dey:2014tfa};
  blue squares -- \cite[Daresbury]{Barber:1981fj} (Panel A);
  red down-triangles -- \cite[CLAS\,2000]{CLAS:2000kid} (Panel B).
  \label{fig:6}}
\end{figure}

The $|t|$ dependence of the $\phi$ photoproduction differential cross section is depicted in Fig.\,\ref{fig:6} for two different but low photon energies.
On $|t|\lesssim 2\,$GeV$^2$, the $\mathbb{P}-$dyn reaction model provides a good description of the data \cite[CLAS\,13]{Dey:2014tfa}, \cite[Daresbury]{Barber:1981fj}, \cite[CLAS\,00]{CLAS:2000kid}.
However, the reaction is not purely diffractive on $|t|\gtrsim 2\,$GeV$^2$: on this domain, one sees structure, likely associated with baryon resonance excitation.
The mismatch between $\mathbb{P}-$am and data on the entire domain depicted highlights the importance of a realistic description of the $\gamma\to\bar{q}q+\mathbb{P}\to V$ transition, Fig.\,\ref{FigPomMechanism}.

\begin{figure*}[t]
\centering
\includegraphics[width=0.24\textwidth]{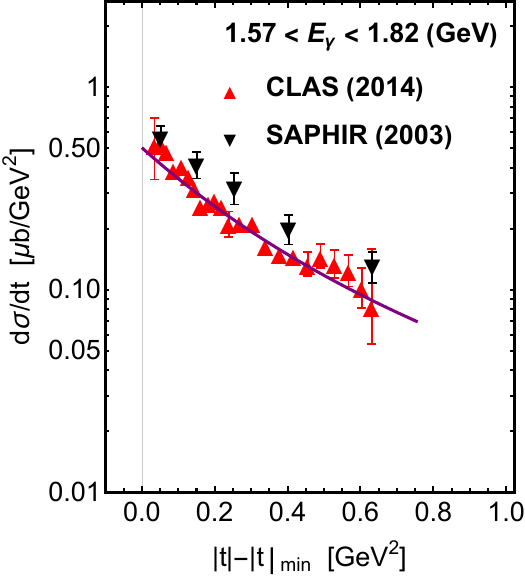}
\includegraphics[width=0.24\textwidth]{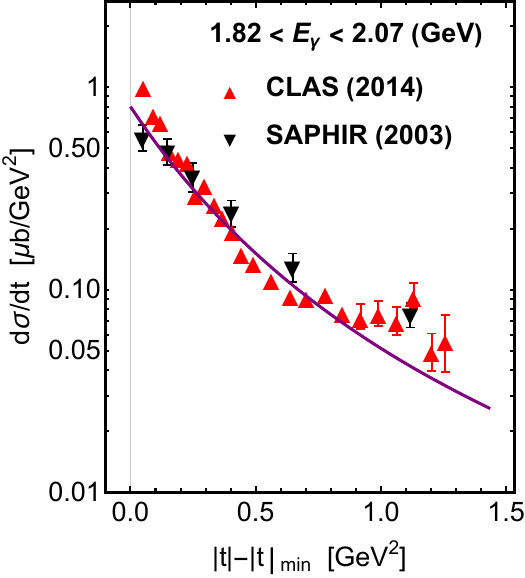}
\includegraphics[width=0.24\textwidth]{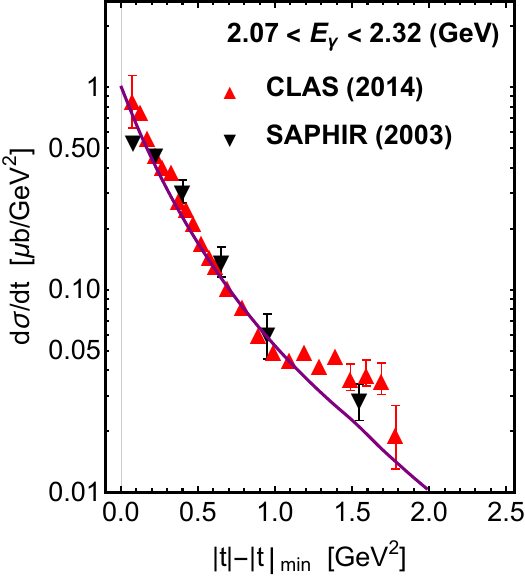}
\includegraphics[width=0.24\textwidth]{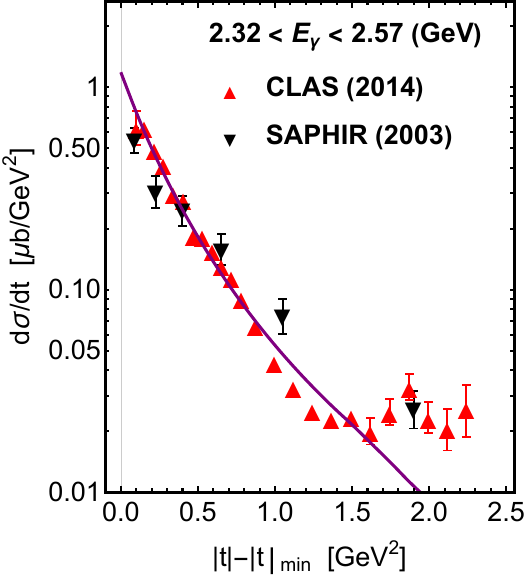}
\caption{Differential cross sections for $\phi$ photoproduction as a function of $|t|-|t|_{\text{min}}$, Eq.\,\eqref{Eqtmin}, at low photon energies, $E_\gamma$.
Solid purple curve  -- $\mathbb{P}-$dyn.
Data: red up-triangles -- \cite[CLAS\,2014]{Dey:2014tfa};
black down-triangles -- \cite[SA\-PHIR]{Barth:2003bq}.
\label{fig:7}}
\end{figure*}

Figures~\ref{fig:7}, \ref{fig:8} compare $\mathbb{P}-$dyn predictions for the $|t|$-dependence of the $\phi$ photoproduction differential cross sections with data available at lower photon energies than those in Fig.\,\ref{fig:6}, \emph{viz}.\ Refs.\,\cite[CLAS\,13]{Dey:2014tfa}, \cite[SA\-PHIR]{Barth:2003bq}, \cite[LEPS]{LEPS:2005hax}.
Regarding Fig.\,\ref{fig:7}, in this case, too, the agreement between our reaction model and data for near-forward scattering is good.  The mismatch at larger $|t|$ was already discussed in connection with Fig.\,\ref{fig:6}.
Concerning Fig.\,\ref{fig:8}, the $\mathbb{P}-$dyn provides a good description of the differential cross section except, perhaps, in the neighbourhood $E_\gamma \simeq 2\,$GeV, whereupon some nondiffractive processes may be contributing somewhat to the production process.

 \begin{figure}[t]
  \centering
  \includegraphics[width=0.475\textwidth]{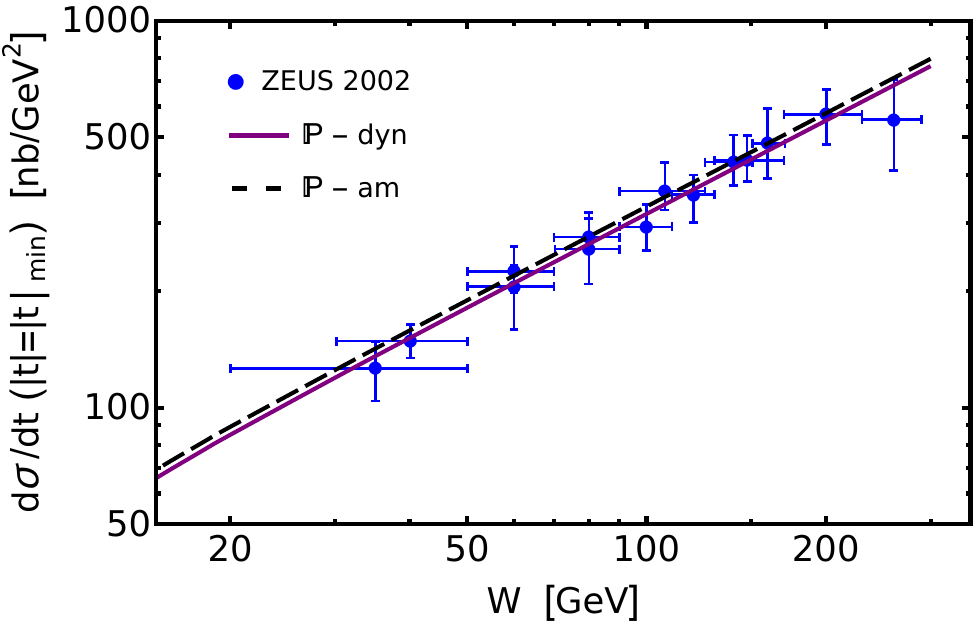}
  \caption{
  Differential cross-section for $\gamma p \to  J/\psi p$ on $|t|/m_p\simeq 0$ (forward).
 Solid purple curve:  $\mathbb{P}-$dyn; dashed black curve -- $\mathbb{P}-$am.
 (Pomeron parameters in Table~\ref{tab:1}\,--\,Row~3.)
Data from Ref.\,\cite[ZEUS\,2002]{ZEUS:2002wfj}.
  \label{fig:10}}
\end{figure}

\subsection{$J/\psi$ photoproduction}
\label{subsecJpsi}
The reaction model characterised by the sketch in Fig.\,\ref{FigPomMechanism}, $\mathbb P-$dyn, was applied to $J/\psi$ photoproduction in Ref.\,\cite{Tang:2024pky}.  Herein, for completeness, we recapitulate and somewhat extend the results described in that study.
Of course, the $J/\psi$ contains no valence degrees of freedom in common with the proton; so, it has long been widely held \cite{Krein:2017usp, Lee:2022ymp} that the dominant mechanisms underlying $J/\psi$ photoproduction must involve some manifestations of gluon physics within the target proton and, perhaps, the incipient vector meson, and/or the exchange of (perhaps infinitely many, correlated) gluons between the partonic constituents of each.  (This latter is a Pomeron-like effect.)

The $J/\psi$ process is topical because some have argued \cite{Kharzeev:1995ij} that there is a connection between near-thres\-hold $J/\psi$ photoproduction and the in-proton expectation value of the QCD trace anomaly; hence, insights into the character of emergent hadron mass (EHM) \cite{Roberts:2016vyn, Krein:2020yor, Roberts:2021nhw, Binosi:2022djx, Ding:2022ows, Ferreira:2023fva, Salme:2022eoy, Carman:2023zke, Achenbach:2025kfx}.
This possibility has served as a motivation for new high-energy, high-luminosity accelerator facilities \cite{Chen:2020ijn, Anderle:2021wcy, AbdulKhalek:2021gbh}.
However, any trace-anomaly connection now appears remote \cite{Du:2020bqj, Xu:2021mju, Sun:2021pyw, Sakinah:2024cza}.
Notwithstanding these remarks, there is a continuing desire to interpret $\gamma + p \to V + p$ photoproduction data in terms of gluon physics; hence, an elucidation of the underlying reaction mechanism is crucial.

\begin{figure}[t]
  \centering
  \includegraphics[width=0.475\textwidth]{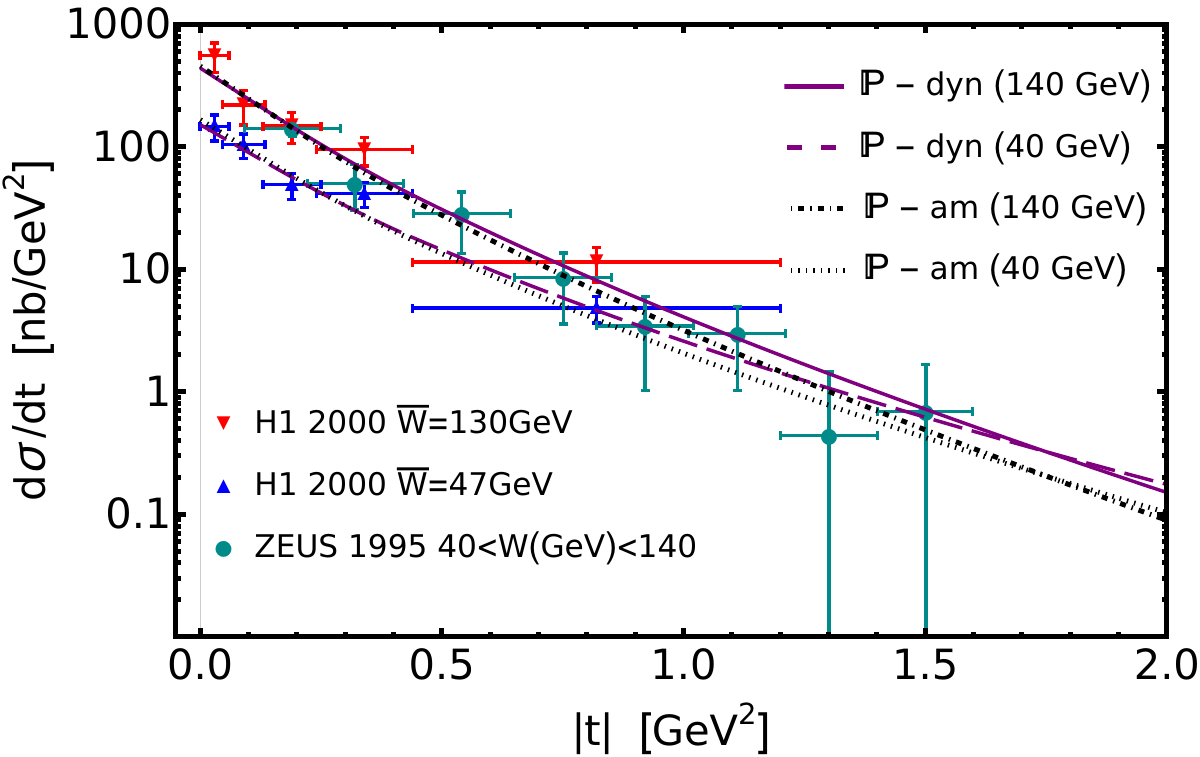}
  \caption{Predicted $|t|$ dependence of differential cross section for $J/\psi$ photoproduction at selected $W$ values well above threshold.
  Solid purple curve -- $\mathbb{P}-$dyn, $W=140\,$GeV;
  dashed purple curve -- $\mathbb{P}-$dyn, $W=40\,$GeV;
  dot-dashed black curve -- $\mathbb{P}-$ am, $W=140\,$GeV;
  dotted black curve -- $\mathbb{P}-$ am, $W=40\,$GeV.
  Data, with $\bar W$ being experimental mean $W$:
  blue up-triangles -- \cite[H1\,2000]{H1:2000kis}; 
  red down-triangles -- \cite[H1\,2000]{H1:2000kis}; 
  cyan circles --  \cite[ZEUS\,1995]{ZEUS:1995kab}. 
  \label{fig:11}}
\end{figure}

As discussed in Sect.\,\ref{sec:24} and in Ref.\,\cite{Tang:2024pky}, the Pomeron--$c$-quark coupling was fixed by requiring a best least-squares fit to the $W$-dependence of the $|t|\approx |t|_{\rm min}$ $J/\psi$ photoproduction differential cross section in Ref.\,\cite[ZEUS 2002]{ZEUS:2002wfj}.  The results are displayed in Fig.\,\ref{fig:10}.
From this point, all $J/\psi$ results are predictions of our reaction model.  Thus, it is significant that, as illustrated in\linebreak Fig.\,\ref{fig:11}, our reaction model provides a good description of the $(W,t)$ dependence of the differential cross sections reported in Ref.\,\cite[H1\,2000]{H1:2000kis}.
Evidently, with increasing $|t|$ high-$W$  data retain some sensitivity to the dynamical quark loop in Fig.\,\ref{FigPomMechanism}: compare $\mathbb P-$dyn with $\mathbb P-$am on $|t| \gtrsim 1\,$GeV$^2$.

\begin{figure}[t]
  \centering
  \includegraphics[width=0.475\textwidth]{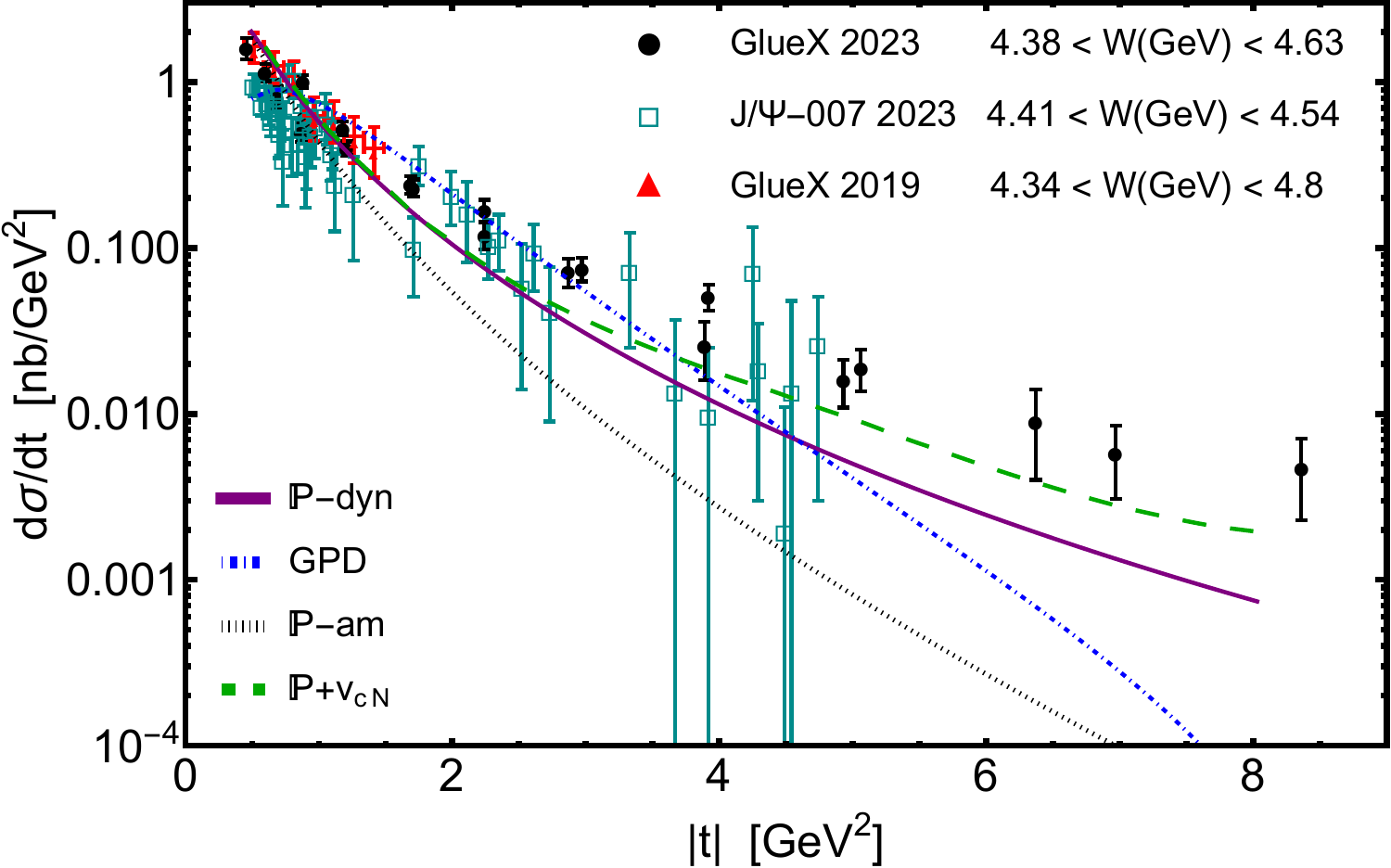}
  \caption{\label{fig:12}
Differential cross-section for $\gamma p \to  J/\psi p$.
Solid purple curve -- $\mathbb{P}-$dyn;
dotted purple curve -- $\mathbb{P}-$am;
dot-dashed blue curve -- GPD model \cite{Guo:2023pqw};
dashed green curve --  Pomeron + $J/\psi N$ scattering model \cite{Sakinah:2024cza}.
(Predictions herein obtained with Pomeron parameters in Table~\ref{tab:1}\,--\,Row~3.)
%
Data:
$4.43< W/{\rm GeV} < 4.8$ -- \cite[GlueX \,2019]{GlueX:2019mkq};
$4.38< W/{\rm GeV} < 4.63$ -- \cite[GlueX\,2023]{GlueX:2023pev};
$4.41< W/{\rm GeV} < 4.54$ -- \cite[$J/\psi$-007]{Duran:2022xag}.
(All theory curves computed using $W=4.5\,$GeV.)}
\end{figure}

Figure~\ref{fig:12} depicts our predictions for the $\gamma p \to  J/\psi p$ differential cross section as a function of $|t|$ at energies not too far above threshold, $W_{\rm th}^{J/\psi}=4.04\,$GeV.
Evidently, on this $W$ domain, the momentum-dependence of the $\gamma\to c\bar c + \mathbb P \to J/\psi$ transition loop in Fig.\,\ref{FigPomMechanism} has a significant influence. 
Again, this is highlighted by the difference between the $\mathbb{P}-$dyn and $\mathbb P\,-$am predictions.

Along with our parameter-free predictions, we also draw the result obtained using a GPD model \cite{Guo:2023pqw}, which involves three parameters and assumes specific, simple forms for the two gluon gravitational form factors: the parameters were fixed via a least-squares fit to the data in Refs.\,\cite{GlueX:2023pev, Duran:2022xag}.
Citing Ref.\,\cite{Tang:2024pky}, quantitatively, the GPD model's description of $J/\psi$-007 data \cite{Duran:2022xag} is better than its match with the GlueX points \cite{GlueX:2023pev}.
In this outcome, there is a signal of overfitting, the probability of which is increased by the result for $d\sigma/dt$ being a concave function on the entire plotted $|t|$ domain, with a global maximum on $|t|\simeq 0.7\,$GeV$^2$.
This is qualitatively inconsistent with the other reaction models, which deliver a differential cross-section that is a monotonically decreasing convex function.

The remaining curve in Fig.\,\ref{fig:12} is the result from Ref.\,\cite{Sakinah:2024cza}, which argues that $J/\psi$ photoproduction should be described by Pomeron exchange augmented by a $c$-quark+nucleon potential, $v_{cN}$, that drives $J/\psi \, N$ scattering and FSIs, with the FSIs dominating near threshold.
This result is similar to our favoured $\mathbb{P}-$dyn prediction on $|t|\lesssim 2.5\,$GeV$^2$ and thereafter becomes harder.  Again citing Ref.\,\cite{Tang:2024pky}, quantitatively, the FSIs in the $\mathbb P +v_{cN}$ reaction model improve the description of GlueX data but degrade that of $J/\psi$-007.
Looking harder, one finds that the most reasonable description of available near-threshold $d\sigma/dt$ data is provided by our $\mathbb{P}-$dyn reaction model: this parameter-free prediction delivers a monotonically decreasing convex differential cross-section and a $\chi^2/$degree-of-freedom that does not vary excessively across the data sets \cite{Tang:2024pky}.

These observations show that whilst modern Poin\-car\'e covariant calculations do indicate that the proton mass radius, $r_m$, is smaller than its charge radius \cite{Yao:2024ixu, Binosi:2026aaa}, owing to the different character of the effective probe \cite{Xu:2023bwv}, one is not justified in drawing any link between the Ref.\,\cite[$J/\psi$-007]{Duran:2022xag} data and a measurement of $r_m$ or, indeed, any other quantity inferred via a GPD-specific interpretation of those or other data.

\begin{figure}[t]
\leftline{\hspace*{0.5em}{\large{\textsf{A}}}}
\vspace*{-2ex}
\centering
 \includegraphics[width=0.46\textwidth]{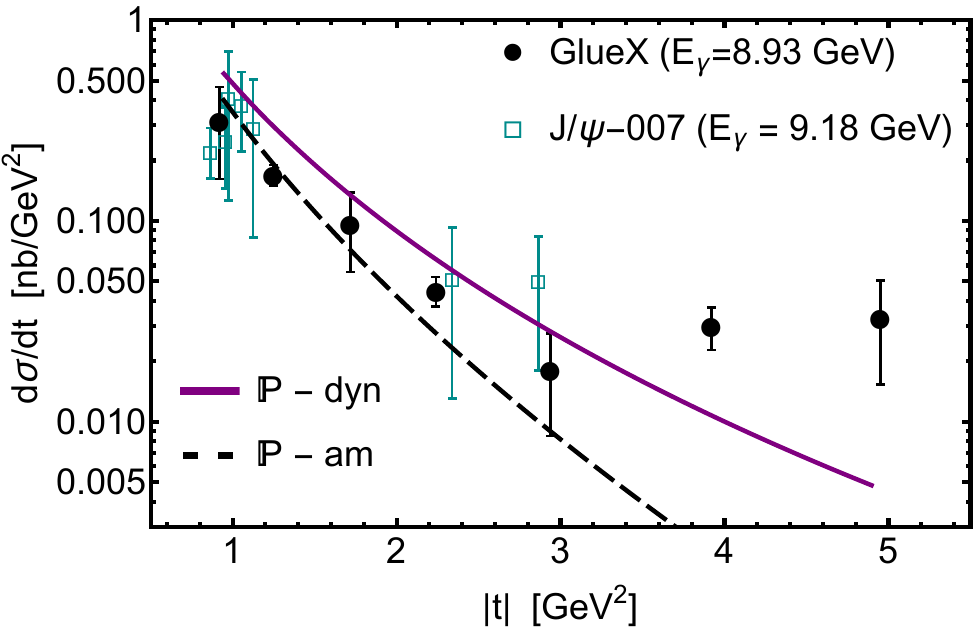}
  \vspace*{2ex}

\leftline{\hspace*{0.5em}{\large{\textsf{B}}}}
\vspace*{-2ex}
\centering
  \includegraphics[width=0.46\textwidth]{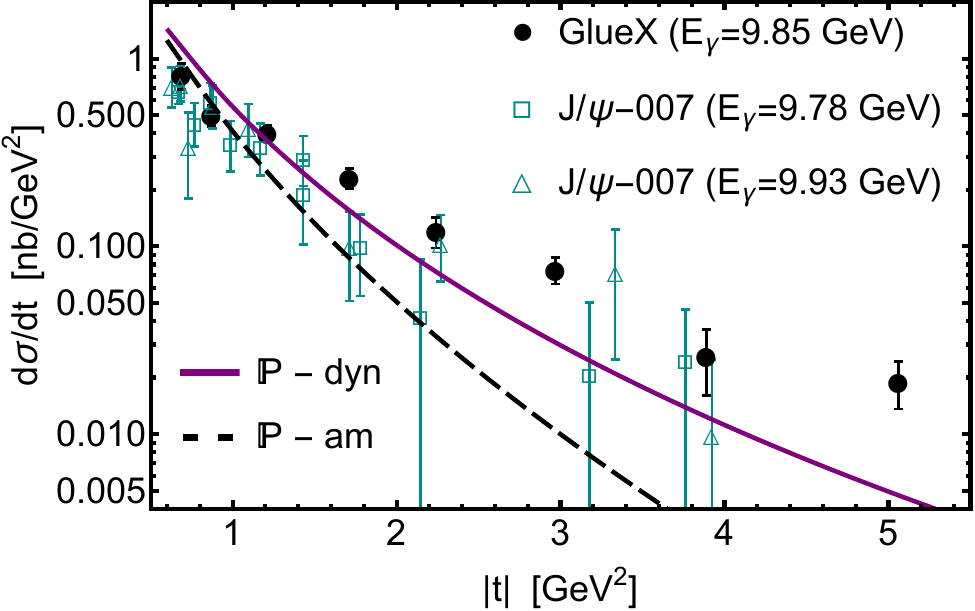}
  \vspace*{2ex}

\leftline{\hspace*{0.5em}{\large{\textsf{C}}}}
\vspace*{-2ex}
\centering
  \includegraphics[width=0.46\textwidth]{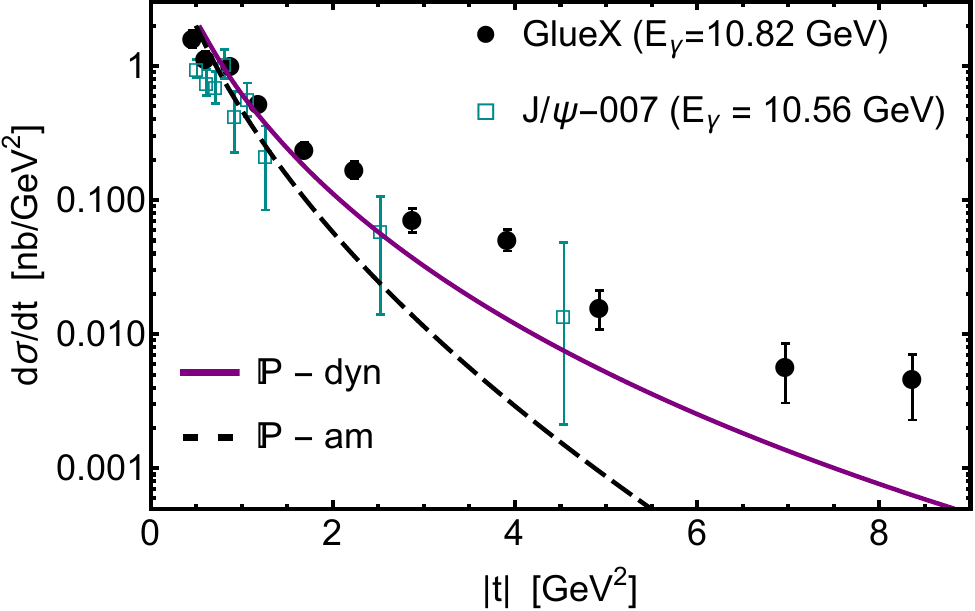}

  \caption{$|t|$ dependence of $J/\psi$ photoproduction differential cross sections near threshold, $W_{\rm th}^{J/\psi} = 4.04\,$GeV.
  Solid purple curve -- $\mathbb{P}-$dyn;
  dotted purple -- $\mathbb{P}-$am.
  The different panels depict, respectively photon energies $E_\gamma = 8.93, 9.85, 10.82 \,$GeV, which correspond to $W = 4.2, 4.4, 4.6\,$GeV. 
  Data: black circles -- \cite[GlueX]{GlueX:2023pev}; cyan squares and triangles -- \cite[$J/\psi$-007]{Duran:2022xag}.
  \label{fig:13}}
\end{figure}


We take a closer look at near-threshold $J/\psi$ photoproduction in Fig.\,\ref{fig:13} by exploring the $(E_\gamma,|t|)$ dependence of the differential cross-section.
Very close to threshold, Fig.\,\ref{fig:13}\,A, the Ref.\,\cite[GlueX]{GlueX:2023pev} data suggest a role for some nondiffractive contributions to the reaction on $|t|\gtrsim 3\,$GeV$^2$.
This possibility may also receive support from the improved match between the Ref.\,\cite{Sakinah:2024cza} (FSI) model and data on that domain.
The domain $|t|\gtrsim 3\,$GeV$^2$ is not reached by the Ref.\,\cite[$J/\psi$-007]{Duran:2022xag} data: on their domain of coverage, those data are entirely consistent with our $\mathbb{P}-$dyn reaction model.
Reviewing Figs.\,\ref{fig:13}\,B, C, it is evident that, as one would expect, the importance of any nondiffractive contributions to the differential cross section diminishes with increasing $E_\gamma$.


\begin{figure}[t]
\leftline{\hspace*{0.5em}{\large{\textsf{A}}}}
\vspace*{-2ex}
\centering
\includegraphics[width=0.475\textwidth]{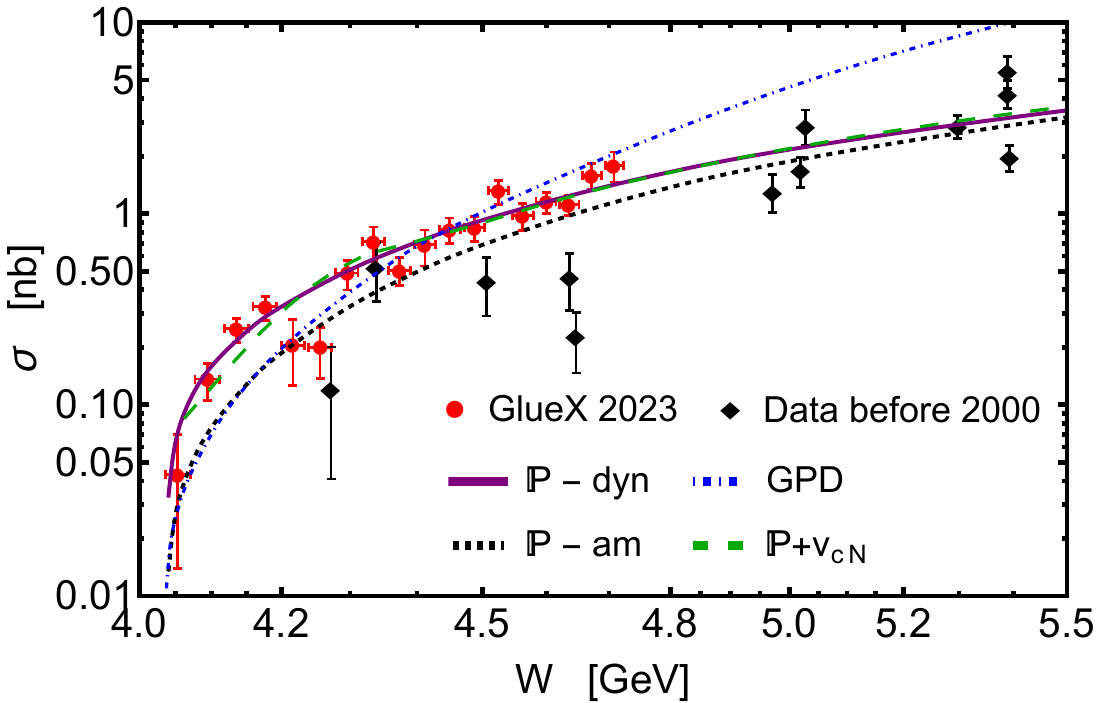}

 \vspace*{2ex}

\leftline{\hspace*{0.5em}{\large{\textsf{B}}}}
\vspace*{-2ex}
\centering
\includegraphics[width=0.475\textwidth]{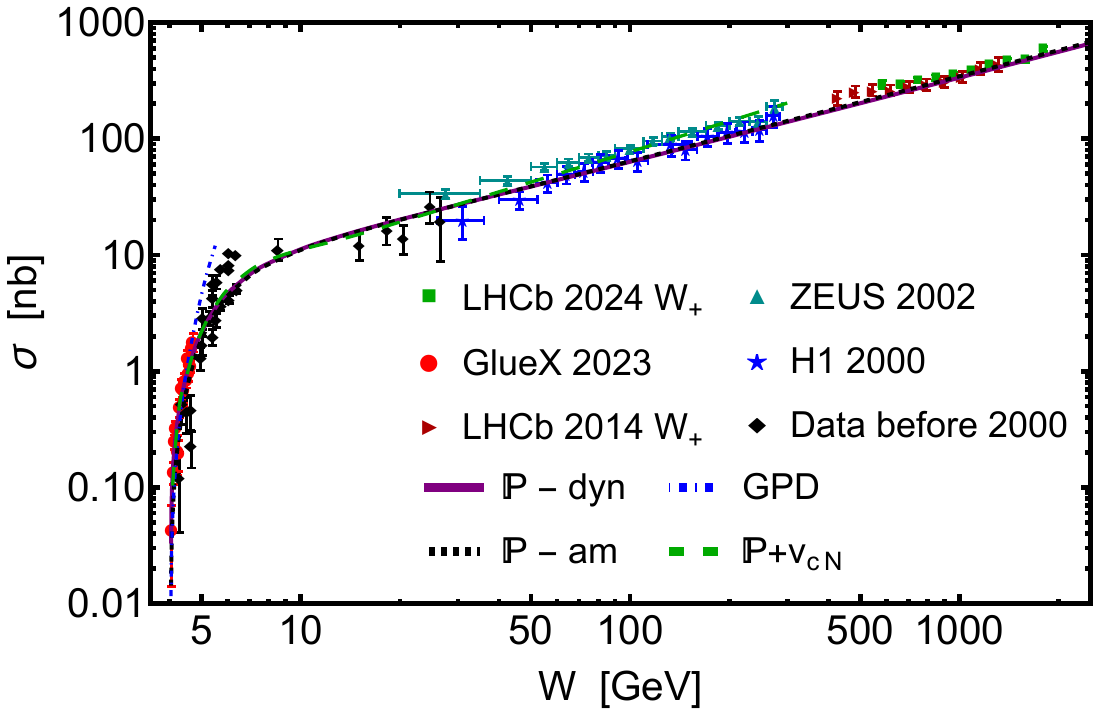}

 \caption{
  $W$ dependence of total $J/\psi$ photoproduction cross section, $W_{\rm th}^{J/\psi} = 4.04\,$GeV.
  {\sf Panel A}. Near-threshold region, $W_{\rm th}^{J/\psi} < W < 5.5\,$GeV.
  {\sf Panel B}. Entire coverage of available data, $W_{\rm th}^{J/\psi} < W < 2\,$TeV.
  Solid purple curve -- $\mathbb{P}-$dyn;
  dashed black -- $\mathbb{P}-$am.
  Data:
  blue stars -- \cite[H1\,2000]{H1:2000kis};
  cyan up-triangles -- \cite[ZEUS \.2002]{ZEUS:2002wfj};
  red right-triangles -- \cite[LHCb\,2014]{LHCb:2014acg};
  red circles -- \cite[GlueX 2023]{GlueX:2023pev};
  green squares --  \cite[LHCb\,2024]{LHCb:2024pcz};
  black diamonds -- pre-2000 \cite{Camerini:1975cy, Gittelman:1975ix, Shambroom:1982qj, ZEUS:1995kab, E687:1993hlm, H1:1996gwv, Amarian:1999pi}.
  \label{fig:14}}
\end{figure}

Our predictions for the $W$ dependence of the total $\gamma p \to J/\psi p$ photoproduction cross section are displayed in Fig.\,\ref{fig:14}.
As explained in Ref.\,\cite{Tang:2024pky}, the $\mathbb{P}-$dyn reaction model provides a good description of all available data
\cite[H1\,2000]{H1:2000kis}, \cite[ZEUS \.2002]{ZEUS:2002wfj}, \cite[LHCb\,2014]{LHCb:2014acg},  \cite[GlueX 2023]{GlueX:2023pev}, \cite[LHCb\,2024]{LHCb:2024pcz}, and pre-2000 \cite{Camerini:1975cy, Gittelman:1975ix, Shambroom:1982qj, ZEUS:1995kab, E687:1993hlm, H1:1996gwv, Amarian:1999pi}, which today covers a domain from threshold to $W\simeq 2\,$TeV.
Near threshold, a realistic description of the $\gamma\to {\mathpzc q}\bar {\mathpzc q} +\mathbb{P}\to V$ transition matrix element in Fig.\,\ref{FigPomMechanism} is crucial in achieving agreement with data: the $\mathbb{P}-$am model fails on this domain, becoming viable only on $W\gtrsim 5\,$GeV.

\begin{figure}[t]
\leftline{\hspace*{0.5em}{\large{\textsf{A}}}}
\vspace*{-2ex}
\centering
 \includegraphics[width=0.46\textwidth]{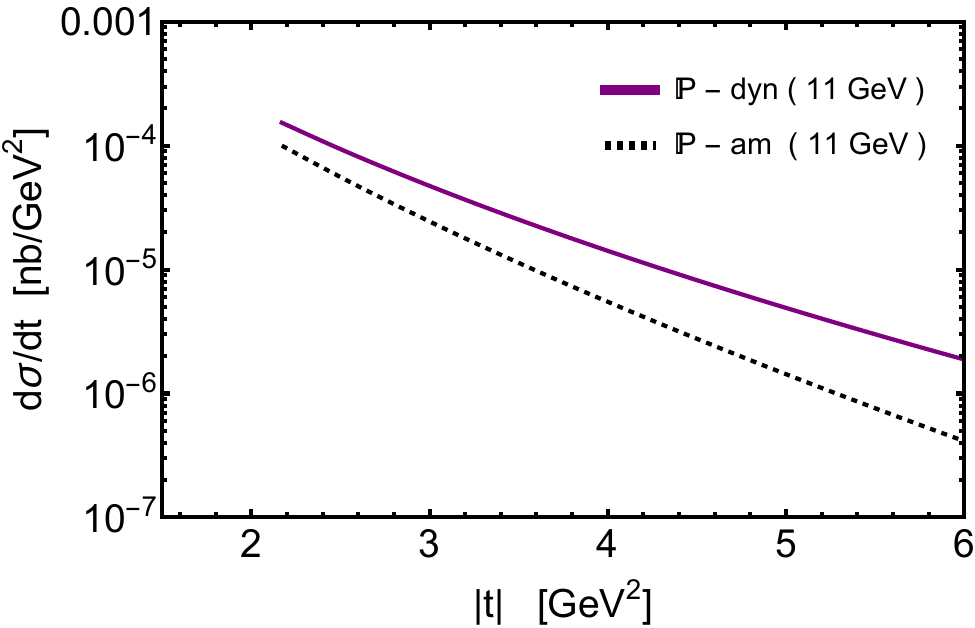}
  \vspace*{2ex}

\leftline{\hspace*{0.5em}{\large{\textsf{B}}}}
\vspace*{-2ex}
\centering
  \includegraphics[width=0.46\textwidth]{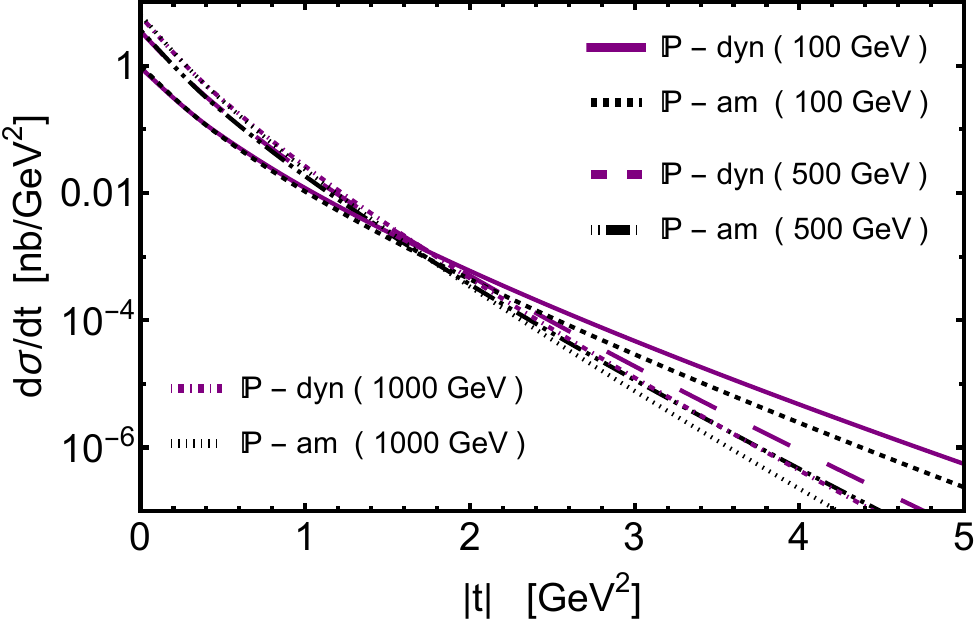}
  \vspace*{2ex}

  \caption{$|t|$ dependence of differential cross section for $\Upsilon$ photoproduction.
   {\sf Panel A}. $W = 11\,$GeV.
   {\sf Panel B}. $\mathbb{P}-$dyn:
   solid purple -- $W = 100\,{\rm GeV}$;
   long-dashed purple -- $500\,$GeV;
   dot-dashed purple  -- $1\,$TeV.
   $\mathbb{P}-$am: dashed black -- $W = 100\,{\rm GeV}$;
   dot-dot-dashed black -- $500\,$GeV;
   dotted black -- $1\,$TeV.
  \label{fig:15}}
\end{figure}

Emulating Ref.\,\cite[Fig.\,6]{Tang:2024pky}, Fig.\,\ref{fig:14} also includes comparisons with other reaction models.
Regarding the \linebreak GPD model, the best-fit curve from Ref.\,\cite{Guo:2023pqw} is drawn: the framework is inapplicable on $W\gtrsim 5\,$GeV; hence, it cannot meet a requirement for providing a uniformly applicable approach.
Furthermore, even on the subdomain of assumed applicability, the description of \cite[GlueX]{GlueX:2019mkq, GlueX:2023pev} data does not match that produced by the $\mathbb{P}-$dyn reaction model, \emph{viz}.\ the approach advocated herein.  These facts further impair attempts to connect near threshold $J/\psi$ photoproduction data with the in-proton gluon GPD.



On the other hand, the reaction model in Ref.\,\cite{Sakinah:2024cza}, delivers a good description of available total cross-section data on the domain $W_{\rm th}^{J/\psi} < W < 300\,$MeV.
In this case, considering also the model's results for the differential cross section -- see Fig.\,\ref{fig:11}, it is the link drawn between such agreement and dominance of FSIs near threshold that is debatable.
The connection relies on a particular choice for the Pomeron trajectory.
If one instead uses the trajectory we have described -- Table~\ref{tab:1}\,-\,Row~3 -- namely, that associated with $\mathbb{P}-$dyn, then an equivalent description of data is possible without calling upon FSIs.

\subsection{$\Upsilon$ photoproduction}
To complete the picture, it remains only to consider $\Upsilon$ photoproduction, the threshold for which is $W_{\rm th}^\Upsilon=10.40\,$GeV.
Today, there are no differential cross section data and total cross sections are only available on $100\,\lesssim W/{\rm GeV} \lesssim 2\,000$.
Owing to the large mass of the valence $b$, $\bar b$ quarks, there are attempts to employ perturbative QCD concepts as the basis for explaining available data; see, \emph{e.g}., Refs.\,\cite{Eskola:2023oos, Penttala:2024hvp, Xie:2025jeg}.
As we shall see, this is by no means mandatory.

Our predictions for the $\gamma p \to  \Upsilon p$ differential cross section are drawn in Fig.\,\ref{fig:15}.
Owing to the large $\Upsilon$ mass, $t_{\rm min} \approx 2.2\,$GeV$^2$ near-threshold.  Naturally, it diminishes with increasing $W$.
Regarding Fig.\,\,\ref{fig:15}\,A, it is apparent that even for $\Upsilon$ photoproduction, the dynamical quark loop in Fig.\,\ref{FigPomMechanism} plays a material role near threshold.  However, as in the other cases considered herein, precise data would be necessary to distinguish between competing reaction mechanisms on this domain.
Turning to Fig.\,\ref{fig:15}\,B, it is evident that, with increasing $W$, the Pomeron trajectory dominates the differential cross section on the diffractive domain.
It is worth stressing here that we use the same Pomeron trajectory for both $J/\psi$ and $\Upsilon$ production.

\begin{figure}[t]
  \centering
  \includegraphics[width=0.475\textwidth]{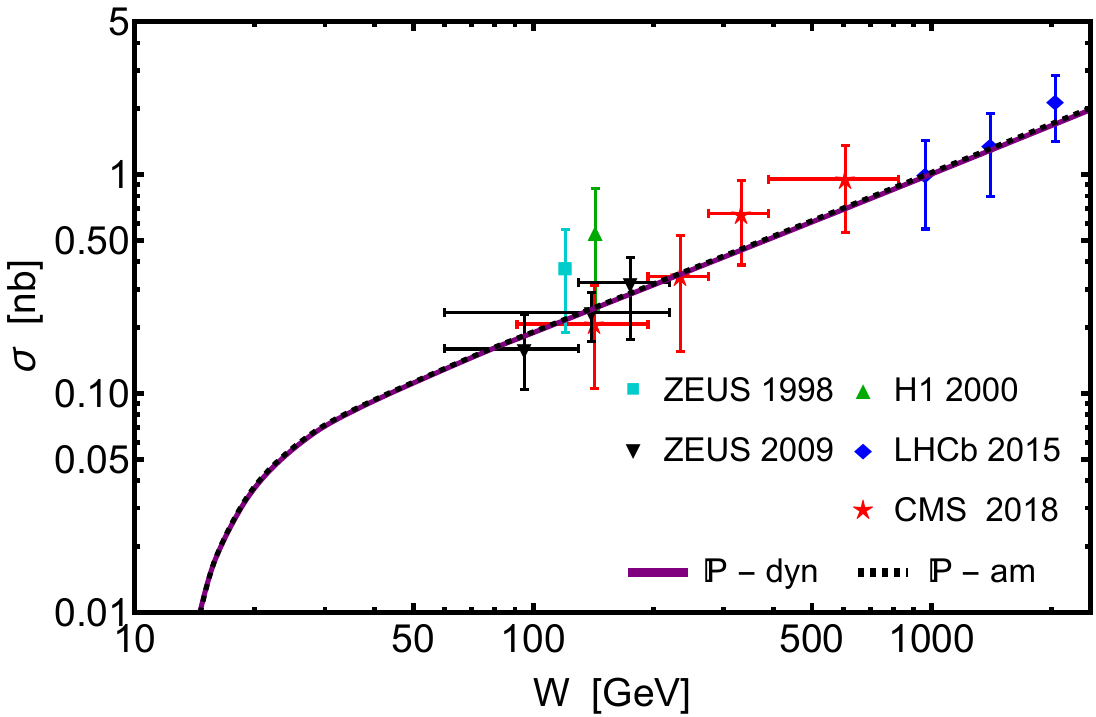}
  \caption{
  $W$ dependence of total cross section for $\Upsilon$ photoproduction, with $W_{\rm th}^\Upsilon=10.40\,$GeV.
  Solid purple curves -- $\mathbb{P}-$dyn;
  dashed black -- $\mathbb{P}-$am.
  Data: cyan squares -- \cite[ZEUS\,1998]{ZEUS:1998cdr};
  green up-triangles -- \cite[H1\,2000]{H1:2000kis};
  black down-triangles -- \cite[ZEUS\,2009]{ZEUS:2009asc};
  blue diamonds -- \cite[LHCb\,2015]{LHCb:2015wlx};
  red stars -- \cite[CMS\,2018]{CMS:2018bbk}.
   \label{fig:16}}
\end{figure}

The total cross section for $\Upsilon$ photoproduction is displayed in Fig.\,\ref{fig:16}.
The $\mathbb{P}-$dyn and $\mathbb{P}-$am reaction models yield equivalent results; so one may conclude that on the domain covered by extant experiments, simple Pomeron exchange is sufficient to describe the data.
Precise data closer to threshold would be valuable.
This may be expected from experiments at future electron ion colliders \cite{Chen:2020ijn, Anderle:2021wcy, AbdulKhalek:2021gbh}.

\begin{figure}[t]
  \centering
  \includegraphics[width=0.475\textwidth]{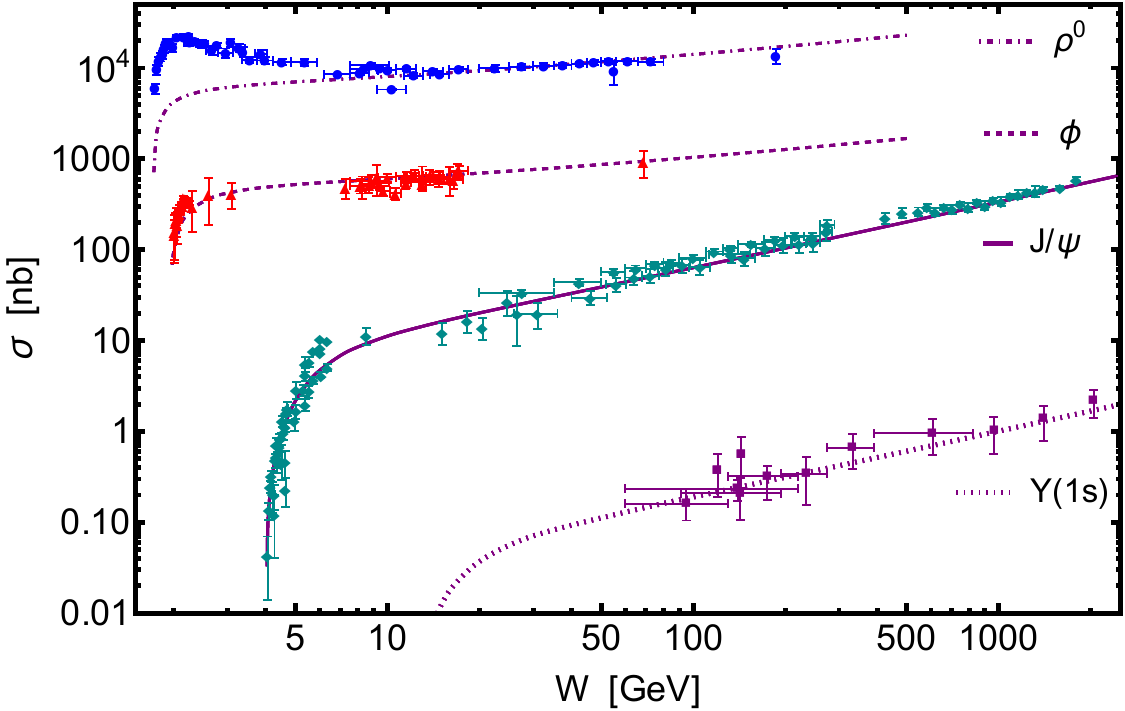}
  \caption{Total cross sections for all vector meson photoproduction reactions considered herein.
  Each curve is the associated $\mathbb{P}-$dyn prediction.
  The data sources are recorded in Figs.\,\ref{fig:4}, \ref{fig:9}, \ref{fig:11}, \ref{fig:16}.
  \label{fig:17}}
\end{figure}

\begin{table}[t]
\centering
\caption{Power-law exponents extracted using Eqs.\,\eqref{PLfits} via least-squares fits to $W$ dependence of $\rho$, $\phi$, $J/\psi$, $\Upsilon$ photoproduction cross sections as calculated using:
{\sf Panel A} -- $\mathbb{P}-$dyn reaction model; and {\sf Panel B} -- $\mathbb{P}-$am.
When employing Eq.\,\eqref{eq:17a}, only $W>W_{\rm min}$ data can be used, \emph{i.e}., data well above threshold.  Herein: $W_{\rm min}^{\rho,\phi} = 5\,$GeV; $W_{\rm min}^{J/\psi} = 10\,$GeV; $W_{\rm min}^{\Upsilon} = 35\,$GeV.
Empirical values taken from Refs.\,\cite{H1:2020lzc, CLAS:2001zwd, H1:2013okq, CMS:2018bbk}.
Somewhat unrefined empirical estimates are also available in Ref.\,\cite[Fig.\,1]{Capua:2012sd}.
\label{tab:3}}
\begin{tabular}{l|ll|ll|l}
\hline
{\sf A}
& \multicolumn{2}{|c|}{Eq.\,\eqref{eq:17a}}
& \multicolumn{2}{|c|}{Eq.\,\eqref{eq:17b}}
& Expt.  \\
\noalign{\smallskip}\hline\noalign{\smallskip}
$V$  & $X$/nb & $\epsilon$ & $\sigma_{P}$/nb & $\epsilon$ & $\epsilon$\\
\noalign{\smallskip}\hline\noalign{\smallskip}
$\rho^{0}$ & $4076$ & 0.27 & 5611 & 0.21 & $0.17 \pm 0.01 ^{+0.04}_{-0.03}$\\
$\phi$ & $234.2$   & 0.26 & 430.0 & 0.20 & $\approx \epsilon_\rho$ \cite{Capua:2012sd, CLAS:2001zwd}\\
$J/\psi$ & $2.136$   & 0.73 & 2.178 & 0.73 & $0.67 \pm 0.03$\\
$\Upsilon$ & $0.0064$  & 0.73 & 0.0066 & 0.73 & $0.77\pm0.14$\\
\hline
\end{tabular}

\vspace*{1ex}

\begin{tabular}{l|ll|ll|l}
\hline
{\sf B}
& \multicolumn{2}{|c|}{Eq.\,\eqref{eq:17a}}
& \multicolumn{2}{|c|}{Eq.\,\eqref{eq:17b}}
& Expt.  \\
\noalign{\smallskip}\hline\noalign{\smallskip}
$V$ & $X$/nb & $\epsilon$ & $\sigma_{P}$/nb & $\epsilon$ & $\epsilon$\\
\noalign{\smallskip}\hline\noalign{\smallskip}
$\rho^{0}$ & $3368$ & 0.32 & 4151 & 0.27 & $0.17 \pm 0.01 ^{+0.04}_{-0.03}$\\
$\phi$ & $268.6$   & 0.30 & 343.7 & 0.25 & $\approx \epsilon_\rho$ \cite{Capua:2012sd, CLAS:2001zwd}\\
$J/\psi$ & $2.099$   & 0.74 & 2.139 & 0.74 & $0.67 \pm 0.03$\\
$\Upsilon$ & $0.0067$  & 0.74 & 0.0066 & 0.73 & $0.77\pm0.14$\\
\hline
\end{tabular}
\end{table}




\section{Results: power-law behavior of vector meson photoproduction cross sections}
\label{SecPower}
Figure~\ref{fig:17} records the total cross sections for all vector meson photoproduction reactions considered herein, depicting $\mathbb{P}-$dyn predictions vs.\ data.
In the analysis of data, it is common to report power-law exponents that characterise the large-$W$ behaviour of such cross sections.
They are typically extracted using one of the following formulae \cite{Donnachie:1992ny, Klein:2016yzr}:
\begin{subequations}
\label{PLfits}
\begin{align}
\sigma(W) &= X  W^{\epsilon} ,\label{eq:17a}\\
\sigma(W) &= \sigma_P  \left[1 - \frac{(m_p + m_V)^2}{W^2} \right]^2  W^{\epsilon} . \label{eq:17b}
\end{align}
\end{subequations}
\emph{NB}. Equation~\eqref{eq:17b} is applicable $\forall W$, whereas Eq.\,\eqref{eq:17a} can only be used on some domain well above threshold, \emph{viz}.\ $W>W_{\rm min} \gg W_{\rm th}$.

For the $\mathbb{P}-$dyn reaction model, values of the fitting parameters are listed in Table~\ref{tab:3}\,A and compared with empirical inferences.
In all cases, the comparison is good when using Eq.\,\eqref{eq:17b}; and Eq.\,\eqref{eq:17a} serves well for heavy mesons.

Employing analogous $\mathbb{P}-$am results from Figs.\,\ref{fig:4}, \ref{fig:9}, \ref{fig:13}, \ref{fig:15}, one obtains the exponents listed in Table~\ref{tab:3}\,B.
In this case, a favourable comparison with empirical inferences is only achieved for heavy mesons, $V=J/\psi, \Upsilon$.

\begin{table*}[t]
\centering
\caption{
VMD-based scattering lengths -- Eq.\,\eqref{avmd1} -- obtained from near-threshold $V$-meson photoproduction differential cross sections calculated using the $\mathbb{P}-$dyn and $\mathbb{P}-$am reaction models.
$q_{\rm th}$ is the photon momentum at threshold in the centre-of-mass frame.
The values of $|\alpha_{V p}^{\rm VMD}|$ in the final column were calculated using Eq.\,\eqref{INew} and values of ${\mathpzc a}_1$ inferred from Refs.\,\cite{Strakovsky:2025ews, Strakovsky:2020uqs, Strakovsky:2019bev, Strakovsky:2021vyk}.
Note that for $|\alpha_{\Upsilon p}^{\rm VMD}|$ the value of ${\mathpzc a}_1$ in Eq.\,\eqref{sigmafit} was estimated in Ref.\,\cite{Strakovsky:2021vyk} by using near-threshold pseudodata created using the GPD model in Ref.\,\cite{Guo:2021ibg}, which is an earlier version of the GPD model discussed above \cite{Guo:2023pqw}: the GPD reaction model delivers results that disagree with ours by an order of magnitude.
\label{tab:4}}
\begin{tabular*}
{\hsize}
{
l@{\extracolsep{0ptplus1fil}}
|l@{\extracolsep{0ptplus1fil}}
l@{\extracolsep{0ptplus1fil}}
l@{\extracolsep{0ptplus1fil}}
l@{\extracolsep{0ptplus1fil}}
l@{\extracolsep{0ptplus1fil}}
|l@{\extracolsep{0ptplus1fil}}}
\hline
$V$ & $m_V$/GeV & $q_{th}$/GeV & $\Gamma_{V\to e^+ e^-}$/keV &
$|\alpha_{Vp}^{\rm dyn}|$/fm &  $|\alpha_{Vp}^{\rm am}|$/fm & $|\alpha_{Vp}^{\rm VMD}|$/fm \\\hline
$ \rho^0\ $ & 0.775 & 0.6 & 6.97 & 0.184 & 0.097 & $0.31(04)$ \\ 
$ \phi^0\ $ & 1.019 & 0.75 & 1.27 & 0.123 & 0.069 & $0.090(14)$ \\ 
$ J/\psi\ $ & 3.097 & 1.91 & 5.53 & $6.34\times 10^{-3}$ & $3.89\times 10^{-3}\ $ & $6.4(1.1) \times 10^{-3}$ \\ 
$ \Upsilon\ $ & 9.46 & 5.16 & 1.34 & $0.196\times 10^{-3}$ & $0.062\times 10^{-3}\ $ & $1.7(0.1) \times 10^{-3}$ \\ 
\hline
\end{tabular*}
\end{table*}

In the standard Pomeron-alone framework, the exponent should be \cite{Donnachie:1992ny}:
\begin{equation}
\epsilon \approx 4[\alpha_{\mathbb P}(0) - 1]\,.
\label{slope}
\end{equation}
Consideration of Eqs.\,\eqref{eq:1}, \eqref{eq:7}, \eqref{PomProp} indicates that a better estimate is provided by the following formula:
\begin{equation}
\epsilon(t_{\rm min}(W)) \approx 4[\alpha_{\mathbb P}(t_{\rm min}(W)) - 1]\,,
\label{slope2}
\end{equation}
where $|t_{\rm min}(W)|$ decreases with increasing $W$.
Notwithstanding Eq.\,\eqref{slope2}, phenomenological analyses of data use Eqs.\,\eqref{PLfits}; so return a $W$-independent exponent, \linebreak which must therefore be some mean-$\epsilon$, averaged over the measurement domain.
Working from Table~\ref{tab:1}, Eq.\,\eqref{slope} yields $\epsilon_\rho = \epsilon_\phi = 0.4$ and $\epsilon_{J/\psi} = \epsilon_\Upsilon = 0.8$.
Evidently, considering Table~\ref{tab:3}, whilst Eq.\,\eqref{slope} is a fair approximation for heavy meson data, $V=J/\psi, \Upsilon$,
it delivers a poor estimate for light meson data, $V=\rho, \phi$.


We note that both the $\mathbb{P}-$dyn and $\mathbb{P}-$am reaction models deliver exponents in qualitative agreement with Eqs.\,\eqref{slope}, \eqref{slope2}: since $\alpha_0^{\rho} = \alpha_0^{\phi}$, $\alpha_0^{J/\psi} = \alpha_0^{\Upsilon}$ in Table~\ref{tab:1}, and $|t_{\rm min}|$ is small, then $\epsilon_\rho \approx \epsilon_\phi$ and $\epsilon_{J/\psi}\approx \epsilon_{\Upsilon}$.

On the other hand, whereas there is quantitative agreement between Eq.\,\eqref{slope} and both $\mathbb{P}-$dyn and $\mathbb{P}-$am predictions for $V=J/\psi, \Upsilon$, this is not the case for $V=\rho, \phi$.
An explanation is found in comparing the predictions in Table~\ref{tab:3}\,A with those in Table~\ref{tab:3}\,B.
The $\mathbb{P}-$am model -- see Fig.\,\ref{Pomam} -- ignores the quark + antiquark transition loop in Fig.\,\ref{FigPomMechanism}.  It only keeps some $t$-dependence via Eq.\,\eqref{PomamFV};
hence $\mathbb{P}-$am is closer to a pure Pomeron description.
Consequently, with less additional interaction structure, the $V=\rho, \phi$ $\mathbb{P}-$am exponents are closer to those anticipated from Eq.\,\eqref{slope}.
A small mean value of $| \bar t_{\rm min}| = 0.061\,$GeV$^2 \approx |t_{\rm min}^\rho|/4$ in Eq.\,\eqref{slope2} yields $\epsilon(\bar t_{\rm min}) = 0.32$, matching the value in Table~\ref{tab:3}\,B.

This discussion highlights that structure in the $\gamma \to V$ transition component of the $\gamma p \to V p$ matrix element has an impact on the exponent value.
Naturally, the effects are greater for the $\mathbb{P}-$dyn reaction model and, indeed, are just what is needed to deliver agreement with the empirically deduced exponent for light mesons.
As we have shown above, such corrections diminish at the typically larger $W$ values associated with $V=J/\psi, \Upsilon$ photoproduction.

\section{Results: near-threshold slope parameter / vector meson-proton scattering lengths}
\label{SecSlope}
As noted in Sect.\,\ref{sec:4}, despite its weaknesses \cite{Xu:2021mju}, VMD remains a widely used phenomenological tool; and within that framework, there is a simple relationship between the differential cross section for $\gamma p \to Vp$ and the $V-p$ scattering length, namely, Eq.\,\eqref{eq:19}.
Our prediction for the $\gamma p \to Vp$ differential cross section is given in Eq.\,\eqref{eq:7} and this is the formula we used above to calculate all results.

\begin{figure}[t]
  \centering
  \includegraphics[width=0.475\textwidth]{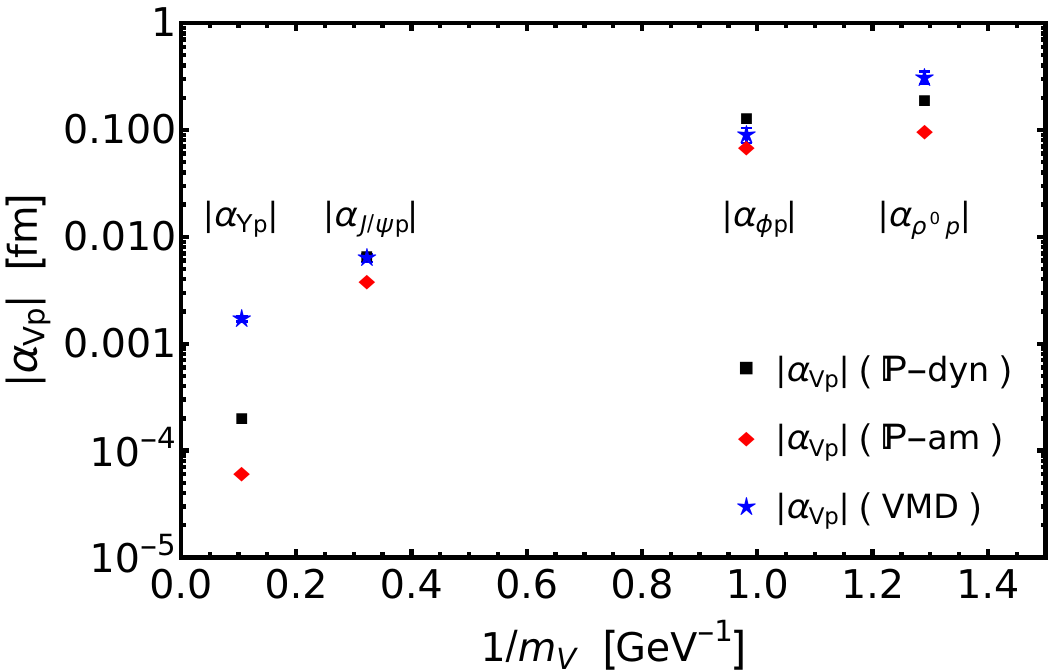}
  \caption{VMD-based scattering lengths -- Eq.\,\eqref{avmd1} -- calculated from near-threshold vector meson photoproduction differential cross sections.  Pictorial representation of the results listed in Table~\ref{tab:4}.
  Legend.
  $\mathbb{P}-$dyn -- black boxes;
  $\mathbb{P}-$am -- red diamonds;
  empirical inferrences -- blue stars.
  %
  \label{fig:18}}
\end{figure}

Here, for the sake of a comparison between our predictions and the model-specific VMD-values of meson-proton scattering lengths inferred in Refs.\,\cite{Strakovsky:2019bev, Strakovsky:2020uqs, Strakovsky:2021vyk, Strakovsky:2025ews}, we equate our near-threshold results with Eq.\,\eqref{eq:19} to arrive at an estimate of the slope parameter $|\alpha_{Vp}|$ via Eq.\,\eqref{avmd1}.
The results are listed in Table~\ref{tab:4} and plotted in Fig.\,\ref{fig:18}.
The differences between $\mathbb{P}-$dyn and $\mathbb{P}-$am predictions again highlight the importance of the dynamical quark loop -- Fig.\,\ref{FigPomMechanism} -- near threshold.
It is worth recording the following pattern in our predictions:
\begin{equation}
 |\alpha_{\Upsilon p}| < |\alpha_{J/\psi p}| \ll |\alpha_{\phi p}| < |\alpha_{\rho^{0} p}|\,.
 \label{eq:21}
\end{equation}
The $\mathbb{P}-$dyn results may be described by
\begin{equation}
|\alpha_{V p}| \approx {\mathpzc a}_0 /m_V^2 , \quad {\mathpzc a}_0 = (0.34\,{\rm GeV})^2\,{\rm fm}.
\end{equation}

As observed in Sect.\,\ref{sec:4}, one need not interpret $|\alpha_{Vp}|$ as the $V-p$ scattering length; instead, it can simply be viewed as a near-threshold slope parameter, in which case the value has objective utility as a comparison between theory and experiment.

We note here that we have compared results obtained using Eq.\,\eqref{avmd1} with those obtained from our total cross sections using Eqs.\,\eqref{sigmafit}, \eqref{INew}.  The results are identical so long as one fits the total cross sections very near threshold, \emph{viz}.\ on $|\vec{P}| \in [0,0.1]\,$GeV.  If one is forced by (experimental) circumstances to begin further from threshold and use a larger range of momenta, then one cannot be certain that the estimate of ${\mathpzc a}_1$ is reliable.

Regarding $V=\rho^0$, there is qualitative agreement between our near-threshold slope and that associated with available data.  One should not expect better because, as explained in Sect.\,\ref{SSrho}, nondiffractive processes are important for $\rho$-meson photoproduction and the empirically inferred value is based on pseudodata \cite{Strakovsky:2025ews}.

Turning to $\phi$ photoproduction, Sect.\,\ref{SSphi} describes good agreement between our predictions and data, so our value for the $\phi$ slope parameter should be reliable.  Hence, the agreement with data is good, especially when one considers the scatter in available near threshold data; see Figs.\,\ref{fig:5}\,--\,\ref{fig:8}, and the use of such data in Ref.\,\cite{Strakovsky:2020uqs} to infer a value of ${\mathpzc a}_1^\phi$.

Owing to the heightened interest in $J/\psi$ photoproduction, a good deal of modern data is available near threshold.  Our prediction for the associated slope parameter agrees well with the value deduced from the new data.

The outlier in Table~\ref{tab:4} is the slope parameter associated with near threshold $\Upsilon$ photoproduction.
In our view, given its success with $J/\psi$ photoproduction, the $\mathbb{P}-$dyn reaction model should be reliable for $\Upsilon$, too.  However, its prediction is an order of magnitude smaller than that inferred from Ref.\,\cite{Strakovsky:2021vyk}.
We judge that the explanation for this mismatch is straightforward.
The estimate in Ref.\,\cite{Strakovsky:2021vyk} is based on pseudodata produced near threshold using the GPD model in Ref.\,\cite{Guo:2021ibg}, which is an earlier version of the GPD model discussed above \cite{Guo:2023pqw}.  As observed in Sect.\,\ref{subsecJpsi} and Ref.\,\cite{Tang:2024pky}, the GPD model provides a poor representation of near-threshold $J/\psi$ data; so one should not expect it to be reliable for $\Upsilon$ photoproduction.
Data from future electron ion collider experiments may be expected to resolve this issue.  Meanwhile, we present our prediction as a solidly benchmarked estimate.

\section{Summary and perspective}
\label{sec:6}
We described and employed a reaction model, $\mathbb P-$dyn, for $\gamma + p \to V + p$ photoproduction, $V=\rho^0, \phi, J/\psi, \Upsilon$, that exposes the quark+antiquark content of the dressed photon in making the transition $\gamma\to {\mathpzc q} \bar{\mathpzc q} + \mathbb P \to V$, where ${\mathpzc q}$ depends on $V$, and couples the intermediate ${\mathpzc q} \bar{\mathpzc q}$ system to the proton's valence quarks via Pomeron exchange [Fig.\,\ref{FigPomMechanism}].
The parameters in this model are those characterising the Pomeron trajectories for light (two) and heavy mesons (two) and the Pomeron-quark couplings (four).
With those parameters fitted to high-$W$ data [Sect.\,\ref{sec:24}], $\mathbb P-$dyn provides a uniformly good description of a large number of differential and total cross-sections for $V$-meson photoproduction from the proton on the entire kinematic ranges measured to date [Sect.\,\ref{sec:5}].
In all cases, the quality of the description is as good as or better than that provided by alternative models.
The quark loop in Fig.\,\ref{FigPomMechanism} plays a key role in the success of the $\mathbb P-$dyn reaction description.

There is one caveat, however.
Namely, as one moves away from the forward limit, nondiffractive mechanisms can play a role in the photoproduction process.
Since the $\rho^0$ contains valence degrees of freedom in common with the proton, that is most apparent for $\rho^0$ production.
In this case, the $\mathbb P-$dyn reaction model is insufficient on a domain that extends from threshold, $W_{\rm th}^\rho$, to $W \approx 5\,$GeV, where $W$ is the centre of mass energy [Fig.\,\ref{fig:4}].

Working with $\mathbb P-$dyn predictions, we extracted the power-law exponents that are often used empirically to characterise the large-$W$ behaviour of total cross sections for $\gamma + p \to V + p$ photoproduction [Sect.\,\ref{SecPower}].  Our results agree well with phenomenological inferences from data.  For light mesons, the exponents do not take the values suggested by simple Pomeron pictures.  Again, the quark loop in Fig.\,\ref{FigPomMechanism} plays a key role in these outcomes.

We also used the $\mathbb P-$dyn predictions to extract near-threshold slope parameters that can be used to characterise the differential cross sections for $\gamma + p \to V + p$ photoproduction [Sect.\,\ref{SecSlope}].
Owing to the role of nondiffractive processes, we judge that the $\rho^0$ value is underestimated.
On the other hand, the predictions for $V=\phi, J/\psi, \Upsilon$ are solidly benchmarked; hence, should serve as realistic estimates.

Acknowledging the $\rho$-meson caveat, our $\mathbb P-$dyn reaction model delivers a uniformly good, unified  description of forward $V=\rho^0, \phi, J/\psi, \Upsilon$ meson photoproduction from relevant thresholds to very high energies.
A key to this success is a realistic nonperturbative description of the $\gamma \to V$ transition, which is especially important near threshold but also plays a role at higher energies.
Apart from the interaction dynamics expressed in treating that loop, the Pomeron provides the only link to ``gluon'' physics in the reaction model and, so far is now known, that mechanism is unrelated to in-proton gluon distributions.
Hence, at present, any attempt to constrain such distributions using near-threshold $\gamma + p \to V + p$ photoproduction data is unjustified.
The desire to make such interpretations should be restrained whilst development of reaction models continues and higher precision data are accumulated.
One may anticipate that subsequent mutual feedback could lead to a well constrained phenomenology which may be employed in understanding $\gamma + p \to V + p$ processes and forging a tight link between the associated data and proton properties.

\begin{CJK*}{UTF8}{gbsn}
\begin{acknowledgements}
We are grateful to
T.-S.\,H.\ Lee, V.\,I.\ Mokeev, I.\ Strakovsky and J.-J.\ Wu for discussions and suggestions.
%
Work supported by:
National Natural Science Foundation of China, grant no.\ 12135007.
\end{acknowledgements}
\end{CJK*}

\begin{small}

\noindent\textbf{Data Availability Statement} Data will be made available on reasonable request.  [Authors' comment: All information necessary to reproduce the results described herein is contained in the material presented above.]
\medskip

\noindent\textbf{Code Availability Statement} Code/software will be made available
on reasonable request. [Authors' comment: No additional remarks.]

\end{small}

\appendix

\section{Tensor structure of the $\gamma \to V$ transition}
\label{App1}
The essentially dynamical component of the reaction model, \emph{viz}.\ the
$\gamma\to {\mathpzc q} \bar {\mathpzc q} + \mathbb P \to V$ transition matrix element, has the tensor structure apparent in Eq.\,\eqref{Eqtmunualpha}.  The tensors take the following form:
{\allowdisplaybreaks
\begin{subequations}
\begin{align}
\tau^{1}_{\mu\nu\alpha}(q,P) & =q_{\alpha} R_{\mu\nu} \,,\\
%
\tau^{2}_{\mu\nu\alpha}(q,P) & = P_{\alpha} R_{\mu\nu} \,,\\
\tau^{3}_{\mu\nu\alpha}(q,P) &= q_{\alpha} q_\nu S_{\nu}\,, \\
\tau^{4}_{\mu\nu\alpha}(q,P)&= P_{\alpha} q_{\mu} S_\nu \,, \\%
\tau^{5}_{\mu\nu\alpha}(q,P)&= q_{\alpha} P_{\mu} S_\nu \,, \\%
\tau^{6}_{\mu\nu\alpha}(q,P)&=P_{\mu} T_{\alpha\nu}\,, \\%
\tau^{7}_{\mu\nu\alpha}(q,P)&=q_{\mu} T_{\alpha\nu} \,, \\%
\tau^{8}_{\mu\nu\alpha}(q,P)&=\delta_{\mu\alpha} S_{\nu} \,, \\%
\tau^{9}_{\mu\nu\alpha}(q,P)&= P_\alpha P_\mu S_{\nu}\,,
\end{align}
\end{subequations}
where
\begin{subequations}
\label{EqTransverse}
\begin{align}
 R_{\mu\nu}& = \delta_{\mu\nu}-P_{\mu}P_{\nu}/P^{2} \,,\\
 S_{\nu}& = q_{\nu}-P_{\nu}P\cdot q/P^{2}\,.
\end{align}
\end{subequations}
}


\end{document}